\documentclass[journal]{IEEEtran}
\usepackage{cite}

\ifCLASSINFOpdf
   \usepackage[pdftex]{graphicx}
   \graphicspath{{figures/}}
   \DeclareGraphicsExtensions{.pdf,.jpeg,.png}
\else
\fi
\usepackage{amsmath}
\usepackage{algorithm}
\usepackage{algorithmic}

\usepackage{stfloats}
\usepackage{url}

\usepackage{amssymb,mathrsfs,bbm}

\DeclareMathOperator{\diag}{diag}
\DeclareMathOperator{\blkdiag}{blkdiag}
\DeclareMathOperator{\prox}{prox}
\DeclareMathOperator{\sign}{sign}

\begin{document}
%
\title{Determined BSS Based on Time-frequency Masking and Its Application to Harmonic Vector Analysis}
%
%
%

\author{
        Kohei~Yatabe,~\IEEEmembership{Member,~IEEE,}
        and~Daichi~Kitamura,~\IEEEmembership{Member,~IEEE}
\thanks{K. Yatabe is with the Department of Intermedia Art and Science, Waseda University, Tokyo 169-8555, Japan (e-mail: k.yatabe@asagi.waseda.jp ).}
\thanks{D. Kitamura is with the Department of Electrical and Computer Engineering, National Institute of Technology, Kagawa College, Kagawa, Japan.}
\thanks{Manuscript received XXXX XX, 20XX; revised XXXX XX, 20XX.}}

%
%

\markboth{IEEE/ACM Transactions on Audio, Speech, and Language Processing,~Vol.~XX, 202X}%
{Yatabe \MakeLowercase{and} Kitamura: Determined BSS Based on Time-frequency Masking and Its Application to Harmonic Vector Analysis}
%



\maketitle

\begin{abstract}
This paper proposes harmonic vector analysis (HVA) based on a general algorithmic framework of audio blind source separation (BSS) that is also presented in this paper.
BSS for a convolutive audio mixture is usually performed by multichannel linear filtering when the numbers of microphones and sources are equal (determined situation).
This paper addresses such determined BSS based on batch processing.
To estimate the demixing filters, effective modeling of the source signals is important.
One successful example is independent vector analysis (IVA) that models the signals via co-occurrence among the frequency components in each source.
To give more freedom to the source modeling, a general framework of determined BSS is presented in this paper.
It is based on the plug-and-play scheme using a primal-dual splitting algorithm and enables us to model the source signals implicitly through a time-frequency mask.
By using the proposed framework, determined BSS algorithms can be developed by designing masks that enhance the source signals.
As an example of its application, we propose HVA by defining a time-frequency mask that enhances the harmonic structure of audio signals via sparsity of cepstrum.
The experiments showed that HVA outperforms IVA and independent low-rank matrix analysis (ILRMA) for both speech and music signals.
A MATLAB code is provided along with the paper for a reference.
\end{abstract}


%
\IEEEpeerreviewmaketitle

\section{Introduction}

\IEEEPARstart{B}{lind} source separation (BSS) is a methodology to recover the source signals from multiple mixtures (audio recordings in the case of this paper) without any knowledge about the mixing system.
Let a convolutive mixing process of the signals be approximated in the time-frequency domain as
\begin{equation}
    \mathbf{x}[t,f]\approx \mathbf{A}[f]\mathbf{s}[t,f],
    \label{eq:mixingProcess}
\end{equation}
where $\mathbf{x} = [x_1,x_2,\ldots, x_M]^\mathrm{T}\in\mathbb{C}^M$ is the observed mixtures obtained by $M$ microphones, $\mathbf{s} = [s_1,s_2,\ldots, s_N]^\mathrm{T}\in\mathbb{C}^N$ is the vector of source signals to be recovered,
$N$ is the number of source signals that is assumed to be known,
$\mathbf{A}[f]\in\mathbb{C}^{M\times N}$ is the mixing matrix, and $t=1,\ldots,T$ and $f=1,\ldots,F$ are indices of time and frequency, respectively.
Throughout this paper, each element of  multichannel signals in $\mathbb{C}^{M\times T\times F}$ is denoted by $x_m[t,f]\in\mathbb{C}$, and the $m$th spectrogram is represented as $x_m[\cdot,\cdot]\in\mathbb{C}^{T\times F}$.
The aim of BSS is to recover the unknown source signals, $\mathbf{s}$, only from the mixtures, $\mathbf{x}$.
This paper considers batch processing, i.e., mixtures for all $t$ are given beforehand.
In the determined ($M \!=\! N$) or overdetermined ($M \!>\! N$) situation, the usual strategy for solving the BSS problem is to formulate an estimation problem of finding (or approximating) a demixing matrix, $\mathbf{W}[f]\in\mathbb{C}^{N\times M}$, that is a left inverse of $\mathbf{A}[f]$ (i.e., $\mathbf{W}[f]\mathbf{A}[f]=\mathbf{I}$, where $\mathbf{I}$ is the identity matrix).
Then, the source signals are recovered by multiplication of the estimated demixing matrix as follows:
\begin{equation}
    \mathbf{W}[f]\mathbf{x}[t,f]\approx \mathbf{W}[f]\mathbf{A}[f]\mathbf{s}[t,f] = \mathbf{s}[t,f].
\end{equation}
By reducing the BSS problem into the demixing matrix estimation problem, the difficulty of directly tackling the unknown mixing process in Eq.~\eqref{eq:mixingProcess} is circumvented.
This paper focuses on the above formulation of the (over)determined BSS, where the demixing matrix $\mathbf{W}[f]$ is estimated for all $f$ only from the observed data $\mathbf{x}[t,f]$.

Statistical independence between the source signals is the well-accepted assumption for handling this ill-posed problem \cite{hyvarinen2000,hyvarinen2004independent}.
While there exists several formulations depending on a method to measure the independence, many of them fall into a minimization problem of the following form \cite{hyvarinen2000}:
\begin{equation}
    \underset{(\mathbf{W}[f])_{f=1}^F}{\text{Minimize}}\;\;\sum_{n=1}^N\mathcal{P}_n(\mathbf{W}[f]\mathbf{x}[t,f])\,-\sum_{f=1}^F\log|\!\det(\mathbf{W}[f])|,
    \label{eq:logDetProblem}
\end{equation}
where the log-determinant term is obtained from either maximum likelihood (ML) estimation or minimizing mutual information \cite{hyvarinen2000}, and $\mathcal{P}_n$ is a real-valued function corresponding to the model of the $n$th source (in the case of ML estimation, $C\exp(-\mathcal{P}_n(\cdot))$ corresponds to the density function of the $n$th source).
For example, with some constant $C$, the $\ell_1$ norm,
\begin{equation}
    \mathcal{P}_n(\mathbf{y}[t,f]) = C\left\|y_n[\cdot,\cdot]\right\|_1 = C\sum_{t=1}^T\sum_{f=1}^F\left|y_n[t,f]\right|,
    \label{eq:ell1}
\end{equation}
recovers the frequency-domain independent component analysis (FDICA) \cite{Smaragdis1998,Araki2003,Sawada2004ICASSP,Buchner2005,Saruwatari2006,Ono2010auxFDICA,ConsistentICA} based on the Laplace distribution, and the $\ell_{2,1}$-mixed norm that treats each time segment as the group,
\begin{equation}
    \mathcal{P}_n(\mathbf{y}[t,f]) = C\left\|y_n[\cdot,\cdot]\right\|_{2,1} = C\sum_{t=1}^T\biggl(\sum_{f=1}^F\left|y_n[t,f]\right|^2\biggr)^{\!\!\frac{1}{2}}\!,
    \label{eq:ell21}
\end{equation}
obtains independent vector analysis (IVA) \cite{Hiroe2006,Kim2006,Kim2007,Ono2011auxIVA} with the spherical Laplace distribution.
The state-of-the art method called the independent low-rank matrix analysis (ILRMA) \cite{Kitamura2016,Kitamura2015,KitamuraEUSIPCO2015} can also be interpreted as Eq.~\eqref{eq:logDetProblem} by considering a function that depends on the rank of each spectrogram,
\begin{equation}
    \mathcal{P}_n(\mathbf{y}[t,f]) = C\,\mathcal{D}_R(y_n[\cdot,\cdot]),
    \label{eq:ILRMApenalty}
\end{equation}
where $\mathcal{D}_R(y_n[\cdot,\cdot])$ is the scalar-valued function that measures low-rankness of a power spectrogram based on the Itakura--Saito non-negative matrix factorization (IS-NMF) \cite{CFevotte2009_ISNMF}:
\begin{align}
    &\mathcal{D}_{R}(y_n[\cdot,\cdot])=\\[-2pt]
    &\min_{\varphi_{f,r}^{[n]}\geq0, \psi_{r,t}^{[n]}\geq0}\sum_{t=1}^T\sum_{f=1}^F\biggl(\frac{\left|y_n[t,f]\right|^2}{\sum_{r=1}^R\!\varphi_{f,r}^{[n]}\psi_{r,t}^{[n]}\!}+\log\sum_{r=1}^R\!\varphi_{f,r}^{[n]}\psi_{r,t}^{[n]}\biggr),\nonumber
\end{align}
where $\varphi_{f,r}^{[n]}$ and $\psi_{r,t}^{[n]}$ capture the spectral and activation patterns of $|y_n[t,f]|^2$, respectively, and $R$ represents rank.

The key to success in these BSS methods is to incorporate prior knowledge on the source signals through the penalty function, $\mathcal{P}_n(\cdot)$.
IVA imposes co-occurrence among the frequency components in each source signal by a frequency-grouped measure as in Eq.~\eqref{eq:ell21}.
Such grouping of frequency components assists in resolving the permutation problem \cite{Kurita2000,Murata2001,Sawada2004Trans,Sawada2007} associated with the frequency-wise treatment of the demixing matrices.
ILRMA takes advantage of a more sophisticated source model to achieve the superior performance.
It assumes low-rankness on the power spectrogram of each source so that inter-frequency and inter-frame dependences of a source signal are captured via NMF.
Recent advancement of (over)determined BSS relies on even more sophisticated source models to improve the separation performance, e.g., super-/sub-Gaussian ILRMA \cite{SuperILRMA2018,SubILRMA2019}, tensor-factorization-based ILRMA \cite{YoshiiILRTA2018,IkeshitaIPSTA2018}, and determined BSS methods based on supervised learning of the source models \cite{IDLMA2019,KameokaVAE2019}.
Therefore, seeking a better source model is important for developing a novel and effective BSS method.

However, the mainstream algorithms as auxIVA \cite{Ono2011auxIVA} and ILRMA \cite{Kitamura2016} cannot easily handle a new source model because they are specialized to each model.
These state-of-the-art algorithms are based on the majorization-minimization (MM) principle \cite{MMbook} that requires upper-bounds approximating the objective function for easier minimization.
Since an upper-bound must be specially designed for each objective function, this requirement forces one to derive a new algorithm each time when a source model is newly defined.
Derivation of the upper-bound is usually heuristic, and it might take a lot of time before examining the performance with a new source model.
One reason for such complication is that those BSS methods are based on the explicit source models (i.e., the source models are explicitly defined as an objective function).
Although a framework based on implicit models has potential of resolving this issue, such framework for determined BSS has rarely been investigated in the literature.

In this paper, to realize effortless investigation of a new source model, we present a general algorithmic framework based on an implicit source model defined via a time-frequency mask.
Since the basic principle of the BSS methods is super-Gaussianity, or sparsity, of the source signals in the time-frequency domain, the difference among the determined BSS methods is the way how to impose the sparsity within their separation processes.
In this respect, the techniques developed with sparsity-based signal processing \cite{Kowalski2009mixedNorm,BachSparsity,Malioutov2014logThresh,Chartrand2016nonConvBook,Chartrand2016invProb,Bayram2017WeakConvex}, such as the proximal splitting technique \cite{Combettes2011,Patrikh2014,convexMonotoneBook2011,Burger2016}, should be beneficial to BSS.
By applying one of the proximal algorithms called primal-dual splitting (PDS) algorithm \cite{Komodakis2015}, the determined BSS problem in Eq.~\eqref{eq:logDetProblem} is handled in a unified manner (Section~\ref{sec:proposedAlgorithms}).
Then, the algorithm is heuristically extended by incorporating a general time-frequency masking method (Section~\ref{sec:generalTFmasking}).
This kind of heuristic extension is called the plug-and-play scheme whose effectiveness has been confirmed in several applications  \cite{Venkatakrishnan2013plugPlay,Chan2017,Ono2017,DeGLIconf,DeGLIjournal}.
The resulted algorithm offers tremendous flexibility into determined BSS because any masking method can be utilized to estimate the demixing matrix, even if the corresponding source model cannot be explicitly written as a formula.

As an application of the general algorithm, we propose a novel BSS method termed \textit{harmonic vector analysis (HVA)}.
To model the source signals, HVA focuses on the harmonic structure of audio signals as a cue for separation.
By considering sparsity of the cepstrum coefficients, the co-occurrence of the harmonic components is captured.
Then, HVA constructs a Wiener-like mask so that the separated signals in each iteration become more exclusive and unmixed.
HVA has the properties of both IVA and ILRMA because HVA can consider the spectral pattern of audio signals as ILRMA while it independently treats each time segment as IVA.
The experimental results showed that the proposed HVA can achieve the state-of-the-art performance for both speech and music signals.

\subsection{Contribution and Outline}

This paper is an extension of the preliminary versions published in the conference proceedings \cite{YatabeICASSP2018,YatabeICASSP2019}.
The contribution of this paper can be summarized as follows:
\begin{itemize}
    \item
    unified and detailed presentation of the separately introduced algorithms \cite{YatabeICASSP2018,YatabeICASSP2019};
    \item
    new extensive experiments for investigating parameters and performance of the algorithms;
    \item
    proposal of a new BSS method, HVA, with some new ideas for realizing it, including cepstrum thresholding, non-separable masking, and cosine shrinkage operator;
    \item
    provision of computational procedures and MATLAB code \cite{HVAcode}.
\end{itemize}

The rest of the paper is organized as follows.
The technical contents begin with some brief explanation of the proximal algorithm and the proximity operator in Section~\ref{sec:preliminaries}.
Then, their application to the determined BSS problem in Eq.~\eqref{eq:logDetProblem} is presented in Section~\ref{sec:proposedAlgorithms}, and its heuristic extension based on time-frequency masking is explained in Section~\ref{sec:generalTFmasking}.
After HVA is proposed in Section~\ref{sec:HVA}, they are experimentally evaluated in Section~\ref{sec:experiment}.
Finally, the paper is concluded in Section~\ref{sec:conclusion}.

\section{Preliminaries}
\label{sec:preliminaries}

\subsection{Primal-dual Splitting (PDS) Algorithm}
\label{sec:PDSexplain}

In this paper, one of the PDS algorithms is adopted for splitting the first and second terms in Eq.~\eqref{eq:logDetProblem}.
At first, let us briefly summarize it in the usual setting.
The PDS algorithm can handle the following general minimization problem:
\begin{equation}
    \underset{\mathbf{w}}{\text{Minimize}}\;\;g(\mathbf{w})+h(\mathbf{L}\mathbf{w}),
    \label{eq:generalProb}
\end{equation}
where $\mathbf{w}$ is the vector to be optimized, $g$ and $h$ are proper lower-semicontinuous convex functions, and $\mathbf{L}$ is a bounded linear operator.
Here, both $g$ and $h$ can be non-differentiable, and hence a gradient-based optimization method may not be applicable for solving it.

When $h$ is non-differentiable in particular, its composition with $\mathbf{L}$ makes the problem difficult.
To handle such difficulty, the associated dual problem is also considered \cite{convexMonotoneBook2011}:
\begin{equation}
    \underset{\mathbf{y}}{\text{Minimize}}\;\;g^\ast(-\mathbf{L}^\ast\mathbf{y})+h^\ast(\mathbf{y}),
    \label{eq:defDualProb}
\end{equation}
where $\mathbf{y}$ is the dual variable, $\mathbf{L}^\ast$ is the adjoint of $\mathbf{L}$, and $g^\ast$ is the Fenchel conjugate of $g$.
Note that this paper avoids explicit consideration of the conjugate by Moreau's identity \cite{convexMonotoneBook2011}, 
\begin{equation}
    \prox_{h^\ast}[\mathbf{z}] = \mathbf{z} - \prox_{h}[\mathbf{z}],
\end{equation}
and therefore we leave its details in the reference \cite{convexMonotoneBook2011}%
\footnote{%
Although Moreau's identity holds only when the function $h$ is convex, this identity is essential for our heuristic extension to a non-convex function.
This is because, by definition, the proximity operator of the Fenchel conjugate of a sparsity-inducing non-convex function is useless for processing.
}
(the definition of the proximity operator, $\prox_h$, will be given in the next subsection).
In the dual problem, $h^\ast$ is free from the linear operator, $\mathbf{L}$, whereas $g$ is free from it in the primal problem in Eq.~\eqref{eq:generalProb}.
Thus, by simultaneously solving the primal and dual problems in Eqs.~\eqref{eq:generalProb} and \eqref{eq:defDualProb}, the iterative procedure of the PDS algorithm can circumvent the composition of $\mathbf{L}$ with the objective functions $g$ and $h$ as follows \cite{Komodakis2015}:
\begin{equation}
    \left\lfloor
    \begin{array}{l}
        \widetilde{\mathbf{w}} = \prox_{\mu_1 g}\bigl[\,\mathbf{w}^{[k]}-\mu_1\mu_2 \mathbf{L}^\ast \mathbf{y}^{[k]}\,\bigr]   , \\[4pt]
        \,\mathbf{z}\, = \mathbf{y}^{[k]} + \mathbf{L}(2\widetilde{\mathbf{w}}-\mathbf{w}^{[k]}), \\[4pt]
        \,\widetilde{\mathbf{y}} = \mathbf{z} - \prox_{\frac{1}{\mu_2}h}[\,\mathbf{z}\,], \\[4pt]
        (\mathbf{w}^{[k+1]},\mathbf{y}^{[k+1]}) = \alpha (\widetilde{\mathbf{w}},\widetilde{\mathbf{y}}) + (1-\alpha)(\mathbf{w}^{[k]},\mathbf{y}^{[k]}),
    \end{array}\right.
    \label{eq:PDSalgorithm}
\end{equation}
where $\mathbf{z}$ is a temporary variable introduced for simpler notation, $\mu_1>0$ and $\mu_2>0$ are step sizes, and $2>\alpha>0$ is a parameter that adjusts the speed of convergence [note that the last line of Eq.~\eqref{eq:PDSalgorithm} can be omitted when $\alpha=1$ is chosen].
The above iterative procedure enables full splitting of the optimization problem in Eq.~\eqref{eq:generalProb}.
That is, the objective functions, $g$ and $h$, as well as the linear operator, $\mathbf{L}$, can be calculated independently of each other.
Therefore, changing one of them ($g$, $h$ or $\mathbf{L}$) only requires modification of the corresponding operator while the others are intact.

For guaranteed convergence, the step-size parameters must be chosen to satify the following inequality \cite{Komodakis2015}:%
\footnote{Strictly speaking, some conditions on the problem (such as non-emptiness of the solution set) are necessary for discussing the convergence \cite{convexMonotoneBook2011}.}
\begin{equation}
    \mu_1\mu_2\|\mathbf{L}\|_\text{s}^2\leq1,
    \label{eq:muCond}
\end{equation}
where $\|\cdot\|_\text{s}$ denotes the spectral norm $(\|\mathbf{L}\|_\text{s} = \sigma_1(\mathbf{L}))$, and $\sigma_1(\mathbf{L})$ is the largest singular value of $\mathbf{L}$.
The parameter $\alpha$ can be arbitrarily chosen from $(0,2)$, where $\alpha=1$ is the standard speed, $\alpha>1$ accelerates, and $\alpha<1$ slows down the algorithm.
Note that the above condition is valid only for a convex problem.
The heuristic extension in Section~\ref{sec:generalTFmasking} will remove the theoretical guarantee of the algorithm, and hence empirical convergence must be experimentally investigated for a general problem.
Even so, we will use Eq.~\eqref{eq:muCond} to set the parameters because we empirically found its usefulness.

\subsection{Proximity Operator and Thresholding Operator}

In the above PDS algorithm, the objective functions, $g$ and $h$, are minimized via the \textit{proximity operator} \cite{Patrikh2014}:
\begin{equation}
    \prox_{\mu g}[\mathbf{z}] = \arg\min_{\boldsymbol{\xi}}\Bigl[\,g(\boldsymbol{\xi}) + \frac{1}{2\mu}\left\|\mathbf{z}-\boldsymbol{\xi}\right\|_2^2\,\Bigr],
    \label{eq:defProx}
\end{equation}
where the left-hand side is regarded as an element of the right-hand side that is singleton for convex $g$.
This subproblem is much easier than the original problem in Eq.~\eqref{eq:generalProb}.
Hence, the PDS algorithm splits the original problem into easier subproblems so that the difficulty is alleviated.
The proximity operator is particularly useful for handling a non-differentiable function (e.g., a sparsity-inducing function) or a differentiable function whose gradient is not Lipschitz continuous (e.g., $-\log(\cdot)$).
Since the determined BSS problem in Eq.~\eqref{eq:logDetProblem} consists of such two functions, it seems natural to handle the BSS problem by the proximity operator.

As is well-known, the proximity operators of some sparsity-inducing penalty functions are closely related to the thresholding (or shrinkage) operators.
For example, the proximity operator associated with the $\ell_1$ norm in Eq.~\eqref{eq:ell1} is given by the bin-wise soft-thresholding operator \cite{Patrikh2014},
\begin{equation}
    \bigl(\prox_{\lambda \left\|\cdot\right\|_1}[\mathbf{z}]\bigr)_n[t,f] = \biggl(1-\frac{\lambda}{|z_n[t,f]|}\biggr)_{\!\!+} z_n[t,f],\label{eq:L1prox}
\end{equation}
where $\lambda\geq0$ is the thresholding parameter, $(\cdot)_+=\max\{0,\cdot\}$ is the half-wave rectifier that replaces negative values by zero, and $(\cdot)_n[t,f]$ denotes the $(n,t,f)$th element of the $N\times T\times F$ array.
The proximity operator of the $\ell_{2,1}$-mixed norm in Eq.~\eqref{eq:ell21} is also given by the group-thresholding operator \cite{Patrikh2014},
\begin{equation}
    \bigl(\prox_{\lambda \left\|\cdot\right\|_{2,1}}[\mathbf{z}]\bigr)_{n}[t,f] = \biggl(1-\frac{\lambda}{(\sum_{f=1}^F|z_n[t,f]|^2)^{\!\frac{1}{2}}\!}\biggr)_{\!\!+} z_{n}[t,f].
    \label{eq:L21prox}
\end{equation}
Proximity operators associated with many other sparsity-inducing functions can also be computed as thresholding operators \cite{Kowalski2009mixedNorm}.
While the penalty functions in the above examples are all convex, the proximity operator is also well-defined for some non-convex functions \cite{Bayram2017WeakConvex}, which may be able to induce sparsity more strongly than the convex ones.

\section{PDS Algorithm for Determined BSS}
\label{sec:proposedAlgorithms}

In this section, the PDS algorithm given in Section \ref{sec:PDSexplain} is applied to the general determined BSS problem in Eq.~\eqref{eq:logDetProblem} for obtaining a base algorithm \cite{YatabeICASSP2018}.

\subsection{Reformulation and Vectorization of the BSS Problem}

To apply the PDS algorithm, the BSS problem is reformulated into the form of Eq.~\eqref{eq:generalProb}.
First, to consider the proximity operator, the second term is modified.
Since the determinant of a matrix can be expressed in terms of its singular values as $|\!\det(\mathbf{W}[f])|=\prod_{n=1}^N\sigma_n(\mathbf{W}[f])$, Eq.~\eqref{eq:logDetProblem} can be rewritten as
\begin{equation}
    \underset{(\mathbf{W}[f])_{f=1}^F}{\text{Minimize}}\;\;\,\mathcal{P}(\mathbf{W}[f]\mathbf{x}[t,f])\,-\sum_{f=1}^F\sum_{n=1}^N\log\sigma_n(\mathbf{W}[f]),
    \label{eq:logSigmaProblem}
\end{equation}
where $\sigma_n(\mathbf{W}[f])\geq0$ is the $n$th singular value of $\mathbf{W}[f]$ in descending order.
Note that the penalty function, $\mathcal{P}$, is also slightly generalized by omitting the summation so that it can be a non-separable function.

Next, the optimization variables are vectorized.
All demixing matrices, $(\mathbf{W}[f])_{f=1}^F$, are vectorized and vertically concatenated to construct an $NMF$-dimensional vector $\mathbf{w}$:
\begin{align}
    \mathbf{w} &= [\mathbf{w}[1]^\mathrm{T},\mathbf{w}[2]^\mathrm{T},\ldots,\mathbf{w}[F]^\mathrm{T}]^\mathrm{T}\in\mathbb{C}^{NMF},
    \\
    \mathbf{w}[f] &= \mathsf{vec}(\mathbf{W}[f])\in\mathbb{C}^{NM},
\end{align}
where $\mathsf{vec}$ is the vectorizing operator converting a matrix into the corresponding vector in the row-major numbering scheme,
\begin{equation}
    \mathsf{vec}(\mathbf{W}[f]) \!=\! [W_{\!1,1}[f],\ldots, W_{\!1,M}[f],W_{\!2,1}[f],\ldots, W_{\!N,M}[f]]^\mathrm{T}\!\!\!.
\end{equation}
The linear operator that converts the $f$th part of the vector $\mathbf{w}[f]$ back into the matrix $\mathbf{W}[f]$ is also defined as
\begin{equation}
    \mathsf{mat}(\mathbf{w}[f]) = \mathbf{W}[f]\in\mathbb{C}^{N\times M},
\end{equation}
which indicates that $\mathbf{w}[f] = \mathsf{vec}(\mathsf{mat}(\mathbf{w}[f]))$.
With these notations, Eq.~\eqref{eq:logSigmaProblem} can be expressed as follows:
\begin{equation}
    \underset{\mathbf{w}}{\text{Minimize}}\;\;\,\mathcal{P}(\mathbf{X}\mathbf{w})\,-\sum_{f=1}^F\sum_{n=1}^N\log\sigma_n(\mathsf{mat}(\mathbf{w}[f])),
    \label{eq:vectorProblem}
\end{equation}
where $\mathbf{X}$ is an $NTF\times NMF$ sparse matrix constructed by copying the observed data, $\mathbf{x}[t,f]$, as
\begin{align}
    \mathbf{X} &= \blkdiag(\boldsymbol{\chi}[1],\boldsymbol{\chi}[2],\ldots,\boldsymbol{\chi}[F]) \in\mathbb{C}^{NTF\times NMF}\!, \\
    \boldsymbol{\chi}[f] &= \blkdiag(\chi[f], \chi[f], \ldots,\chi[f]) \in\mathbb{C}^{NT\times NM}, \label{eq:copyX}\\
    \chi[f] &= [\tau_1[f],\tau_2[f],\ldots,\tau_M[f]]\in\mathbb{C}^{T\times M}, \\
    \!\!\tau_m[f] &= [x_m[1,f],x_m[2,f],\ldots,x_m[T,f]]^\mathrm{T}\in\mathbb{C}^{T},
\end{align}
and $\blkdiag$ is the operator constructing a block-diagonal matrix by concatenating inputted matrices diagonally.

Let the second term in Eq.~\eqref{eq:vectorProblem} be shortly denoted by $\mathcal{I}$:
\begin{equation}
    \mathcal{I}(\mathbf{w}) = -\sum_{f=1}^F\sum_{n=1}^N\log\sigma_n(\mathsf{mat}(\mathbf{w}[f])).
    \label{eq:defI(w)}
\end{equation}
Then, Eq.~\eqref{eq:vectorProblem} can be rewritten with a compact notation: 
\begin{equation}
    \underset{\mathbf{w}}{\text{Minimize}}\;\;\mathcal{I}(\mathbf{w}) + \mathcal{P}(\mathbf{X}\mathbf{w}).
    \label{eq:PDS-BSSproblem}
\end{equation}
Since this form is the same as Eq.~\eqref{eq:generalProb}, the PDS algorithm in Eq.~\eqref{eq:PDSalgorithm} can be applied at least as a procedure.

\begin{algorithm}[t]
\caption{PDS-BSS \cite{YatabeICASSP2018}}
\label{alg:PDS-BSS}
\begin{algorithmic}[1]
\STATE \textbf{Input:} $\mathbf{X}$, $\mathbf{w}^{[1]}$, $\mathbf{y}^{[1]}$, $\mu_1$, $\mu_2$, $\alpha$
\STATE \textbf{Output:} $\mathbf{w}^{[K+1]}$
\FOR{$k = 1, \ldots, K$}
\STATE $\widetilde{\mathbf{w}} = \prox_{\mu_1 \mathcal{I}}[\;\mathbf{w}^{[k]}-\mu_1\mu_2 \mathbf{X}^\mathrm{H}\mathbf{y}^{[k]}\;]$
\STATE $\,\mathbf{z}\, = \mathbf{y}^{[k]} + \mathbf{X}(2\widetilde{\mathbf{w}}-\mathbf{w}^{[k]})$
\vspace{1pt}
\STATE $\:\widetilde{\mathbf{y}} = \,\mathbf{z} - \prox_{\frac{1}{\mu_2}\mathcal{P}}[\;\mathbf{z}\;]$
\vspace{-1pt}
\STATE $\,\mathbf{y}^{[k+1]} = \alpha\widetilde{\mathbf{y}}+(1-\alpha)\mathbf{y}^{[k]}$
\STATE $\mathbf{w}^{[k+1]} = \alpha\widetilde{\mathbf{w}}+(1-\alpha)\mathbf{w}^{[k]}$
\ENDFOR
\end{algorithmic}
\end{algorithm}

\subsection{PDS Algorithm for Determined BSS}

Direct application of the PDS algorithm to Eq.~\eqref{eq:PDS-BSSproblem} obtains Algorithm~\ref{alg:PDS-BSS}.
To realize BSS with this algorithm, two proximity operators, $\prox_\mathcal{I}$ and $\prox_\mathcal{P}$, must be evaluated.

It is known that the proximity operator of an orthogonally invariant function can be evaluated by applying the corresponding proximity operator to the singular values of the inputted matrix \cite{Patrikh2014}.
By regarding $-\log\sigma_n$ in Eq.~\eqref{eq:defI(w)} as $-\log|\sigma_n|$, the proximity operator of $\mathcal{I}(\mathbf{w})$ is obtained \cite{convexMonotoneBook2011}:
\begin{equation}
    \prox_{\mu \mathcal{I}}[\mathbf{w}] = [(\prox_{\mu \widetilde{\mathcal{I}}}[\mathbf{w}[1]])^\mathrm{T},\ldots,(\prox_{\mu \widetilde{\mathcal{I}}}[\mathbf{w}[F]])^\mathrm{T}]^\mathrm{T},
\end{equation}
where $\prox_{\mu \widetilde{\mathcal{I}}}$ is the following proximity operator that moderately increases the singular values of $\mathsf{mat}(\mathbf{w}[f])$,
\begin{equation}
    \prox_{\mu \widetilde{\mathcal{I}}}[\mathbf{w}[f]] = \mathsf{vec}(\,\mathbf{U}[f]\,\widetilde{\boldsymbol{\Sigma}}(\mathsf{mat}(\mathbf{w}[f]))\,\mathbf{V}[f]^\mathrm{H}),
    \label{eq:proxI}
\end{equation}
$\mathbf{W}[f]=\mathbf{U}[f]\boldsymbol{\Sigma}[f] \mathbf{V}[f]^\mathrm{H\!}$ is the singular value decomposition of $\mathbf{W}[f]\;(=\mathsf{mat}(\mathbf{w}[f]))$, $\widetilde{\boldsymbol{\Sigma}}(\cdot)$ is the diagonal matrix,
\begin{equation}
    \widetilde{\boldsymbol{\Sigma}}(\mathbf{W}) = \diag(\prox_{-\mu\log}[\sigma_1(\mathbf{W})],\ldots,\prox_{-\mu\log}[\sigma_N(\mathbf{W})]),
\end{equation}
whose diagonal elements comprise the modified singular values given by applying the proximity operator of $-\mu\log$ \cite{convexMonotoneBook2011},
\begin{equation}
    \prox_{-\mu\log}[\,\sigma\,] = \bigl(\,\sigma + \sqrt{\sigma^2+4\mu}\,\bigr)/2,
\end{equation}
and $\diag$ is the operator constructing a diagonal matrix from inputted scalars.
In other words, applying the proximity operator of $-\mu\log$ to each singular value of $\mathbf{W}[f]$ gives $\prox_{\mu\mathcal{I}}[\cdot]$ as shown in Algorithm \ref{alg:prox_I}, where $\mathrm{svd}(\cdot)$ denotes the singular value decomposition.
This operation is numerically stable because it does not excessively magnify $\left\|\mathbf{w}\right\|_2$ in contrast to the MM algorithms \cite{Ono2011auxIVA,Kitamura2016} that involve inversion of the matrices, which may lead to instability.

\begin{algorithm}[t!]
\caption{Computation of $\prox_{\mu_1 \mathcal{I}}[\mathbf{w}]$}
\label{alg:prox_I}
\begin{algorithmic}[1]
\STATE \textbf{Input:} $\mathbf{w}$, $\mu_1$
\STATE \textbf{Output:} $\widetilde{\mathbf{w}}$
\FOR{$f = 1, \ldots, F$}
\STATE $[\mathbf{U}, \boldsymbol{\Sigma}, \mathbf{V}]=\mathrm{svd}(\mathsf{mat}(\mathbf{w}[f]))$
\STATE $\Sigma_{n,n} = (\Sigma_{n,n}+(\Sigma_{n,n}^2+4\mu_1)^\frac{1}{2})/2\quad\forall n$
\STATE $\widetilde{\mathbf{w}}[f] = \mathsf{vec}(\mathbf{U}\boldsymbol{\Sigma}\mathbf{V}^\mathrm{H})$
\ENDFOR
\end{algorithmic}
\end{algorithm}

\begin{algorithm}[t!]
\caption{Computation of $\mathbf{X}\mathbf{w}$}
\label{alg:Xw}
\begin{algorithmic}[1]
\STATE \textbf{Input:} $\mathbf{X}$, $\mathbf{w}$
\STATE \textbf{Output:} $\hat{\mathbf{s}}$
\STATE $\hat{\mathbf{s}}[t,f] = \mathsf{mat}(\mathbf{w}[f])\,\mathbf{x}[t,f]\quad\forall t,f$
\end{algorithmic}
\end{algorithm}

\begin{algorithm}[t!]
\caption{Computation of $\mathbf{X}^\mathrm{H}\mathbf{y}$}
\label{alg:X'y}
\begin{algorithmic}[1]
\STATE \textbf{Input:} $\mathbf{X}$, $\mathbf{y}$
\STATE \textbf{Output:} $\mathbf{w}'$ [whose $(n,m,f)$th element is $W'_{n,m}[f]$]
\STATE $W'_{n,m}[f] = \sum_{t=1}^T y_n[t,f]\,\overline{x_m[t,f]}\quad\forall n,m,f$
\end{algorithmic}
\end{algorithm}

Note that the matrix, $\mathbf{X}$, is defined only for the formulation and is unnecessary for the implementation \cite{HVAcode}.
This is because the matrix-vector multiplications, $\mathbf{X}\mathbf{w}$ and $\mathbf{X}^\mathrm{H}\mathbf{y}$, can be algorithmically computed as shown in Algorithm \ref{alg:Xw} and \ref{alg:X'y}, respectively, where the overline in Algorithm \ref{alg:X'y} denotes complex conjugation.
Since the matrix, $\mathbf{X}$, is given by copying the same components as in Eq.~\eqref{eq:copyX}, avoiding its construction can reduce the required amount of memory.

Algorithm~\ref{alg:PDS-BSS} can be applied to many BSS models by only changing $\prox_{\mathcal{P}/\mu_2}[\cdot]$ in the 6th line.
For example, an algorithm for FDICA is obtained by inserting the soft-thresholding operator in Eq.~\eqref{eq:L1prox}, while that for IVA is obtained by the group-thresholding operator in Eq.~\eqref{eq:L21prox}.
Thus, Algorithm~\ref{alg:PDS-BSS} can be used to test performance of BSS models without effort on modifying the code whenever the proximity operator of $\mathcal{P}$ is computable.
A source model consisting of two or more penalty functions can also be easily handled by this PDS algorithm (see Section 3.4 and Algorithm 2 of \cite{YatabeICASSP2018} for details).

\subsection{Some Notes on Practical Issues}
\label{sec:normalization}


In this paper, as explained in Section \ref{sec:PDSexplain}, the inequality condition in Eq.~\eqref{eq:muCond} is applied for setting $\mu_1$ and $\mu_2$.
To do so, $\|\mathbf{X}\|_\text{s}^2$ must be calculated.
This can be easily done by using an iterative algorithm that computes the largest singular value of $\mathbf{X}$.
Since $\|\mathbf{X}\|_\text{s}^2=\|\mathbf{X}^\mathrm{H}\mathbf{X}\|_\text{s}$, applying an iterative method (e.g., the power method) to $\mathbf{X}^\mathrm{H}\mathbf{X}$ instead of $\mathbf{X}$ can reduce the computational cost.

To make the choice of the parameters simpler, the following normalization is considered in this paper:
\begin{equation}
    \widetilde{\mathbf{X}} = \mathbf{X}/\left\|\mathbf{X}\right\|_\text{s}.
    \label{eq:normalization}
\end{equation}
Then, the rule for choosing the step sizes can be simplified as
\begin{equation}
    \mu_1\mu_2=1.
\end{equation}
Therefore, with the above normalization, the number of parameters can be reduced by setting $\mu_2 = 1/\mu_1$.
Note that a computationally cheaper norm can be used in place of $\left\|\mathbf{X}\right\|_\text{s}$ to upper bound the data matrix, $\|\widetilde{\mathbf{X}}\|_\text{s}^2\leq1$.
Typical choices are $1$-norm, $\infty$-norm, and $\max$ norm because they can be quickly computed by comparison of the elements.

The whitening of the observed data \cite{hyvarinen2004independent} is strongly recommended for Algorithm~\ref{alg:PDS-BSS}.
This is because $\mathbf{w}^{[k]}-\mu_1\mu_2 \mathbf{X}^\mathrm{H}\mathbf{y}^{[k]}$ in the 4th line updates the demixing filter, $\mathbf{w}$, in a step-by-step manner like a gradient descent method (see Section \ref{subsec:intuitiveInterp} for some intuition).
The whitening can act as preconditioning that accelerates the optimization algorithms.
In addition, it can normalize the level of the observed signals, and therefore the whitening makes it easier to set a parameter that depends on the scale of the signals, such as $\lambda$ in Eq.~\eqref{eq:L21prox}.
This paper will consider the whitening as default.

As usual, the scales of the separated signals cannot be uniquely determined.
To align the frequency-wise scales, postprocessing based on the minimal distortion principle, called back projection \cite{Matsuoka2002projBack}, is used in this paper.

\section{General Time-Frequency Masking as a Heuristic Substitute of the Proximity Operator}
\label{sec:generalTFmasking}

In this section, the proximity operator in Algorithm \ref{alg:PDS-BSS} is substituted by a time-frequency-masking function \cite{YatabeICASSP2019}.
This modification enables us to design a variety of BSS methods without derivation of the corresponding algorithms.

\subsection{Generalized Thresholding/Shrinkage Operators}

The proximity operators of several sparsity-inducing penalty functions can be computed analytically as in Eqs.~\eqref{eq:L1prox} and \eqref{eq:L21prox}.
However, this is not the case for many other functions.
Although there are some formulas that allow computation of a proximity operator from already known ones, e.g. \cite{GramfortEEG2013L21+L1},
\begin{equation}
    \prox_{\lambda_1 \left\|\cdot\right\|_{2,1} + \lambda_2 \left\|\cdot\right\|_{1} }[\mathbf{z}] = \prox_{\lambda_1 \left\|\cdot\right\|_{2,1}}[\,\prox_{\lambda_2 \left\|\cdot\right\|_{1} }[\mathbf{z}]\,],
    \label{eq:L1L21prox}
\end{equation}
applicability of such easy-to-use formula is limited to a specific class of penalty functions \cite{Combettes2008sumProx,YuNIPS2013sumProx,Pustelnik2017sumProx}.
Thus, one has to run an additional iterative algorithm to compute a proximity operator that cannot be written in a closed form.
This can be regarded as a trade-off between flexibility and computational efficiency because types of proximity operators that can be written in closed forms do not have much variety.


To circumvent such trade-off, the generalized thresholding focuses on closed-form (or cheaply computable) operators by directly defining a thresholding/shrinkage operator without defining the corresponding penalty function \cite{Kowalski2013social,Chartrand2014smoothHard,Kowalski2014threshRule,Selesnick2014maximalSparse,Bayram2015monotonePenalFun}.
For example, the $p$-shrinkage operator \cite{Chartrand2014smoothHard}, defined as
\begin{equation}
    \bigl(\mathcal{T}_{p}^{\lambda}[\mathbf{z}]\bigr)_n[t,f] = \biggl(1-\frac{\lambda^{2-p}}{|z_n[t,f]|^{2-p}}\biggr)_{\!\!+} z_n[t,f],
    \label{eq:p-shrink}
\end{equation}
corresponds to some penalty function that does not have an explicit formula for general $p$.
That is, this shrinkage operator defines an implicit model that induces sparsity.%
\footnote{An element-wise function $(\mathcal{T}[\mathbf{z}])_n \!= a(|z_n|)\sign(z_n)$ is the proximity operator of some function that may not have closed-form expression. The condition for $\mathcal{T}$ to be a proximity operator is that $a$ is non-decreasing, $a(|z_n|)\to\infty$ as $|z_n|\to\infty$, and $0\leq a(|z_n|)\leq|z_n|$ \cite{Kowalski2014threshRule}. This fact motivated us to consider the time-frequency masking in Section \ref{subsec:defTFmaskingBSS}.}


Another example is one of the social sparsity operators \cite{Kowalski2013social},
\begin{equation}
    \bigl(\mathcal{T}_{h}^{\lambda}[\mathbf{z}]\bigr)_n[t,f] = \biggl(1-\frac{\lambda}{(h\ast|z_n[t,f]|^2)^{\!\frac{1}{2}}\!}\biggr)_{\!\!+} z_{n}[t,f],
    \label{eq:socialShrink}
\end{equation}
where $h\ast$ represents the convolution with a two-dimensional filter kernel $h$, whose elements are non-negative, in the time-frequency domain.
Although its effectiveness has been empirically shown \cite{Kowalski2013social}, this operator is not a proximity operator of some function in general \cite{Gribonval2020}.
That is, the social sparsity operator goes beyond the proximity operator and realizes an efficient algorithm with a flexible implicit signal model.

\begin{figure}[!t]
    \centering
    \includegraphics[width=0.99\columnwidth]{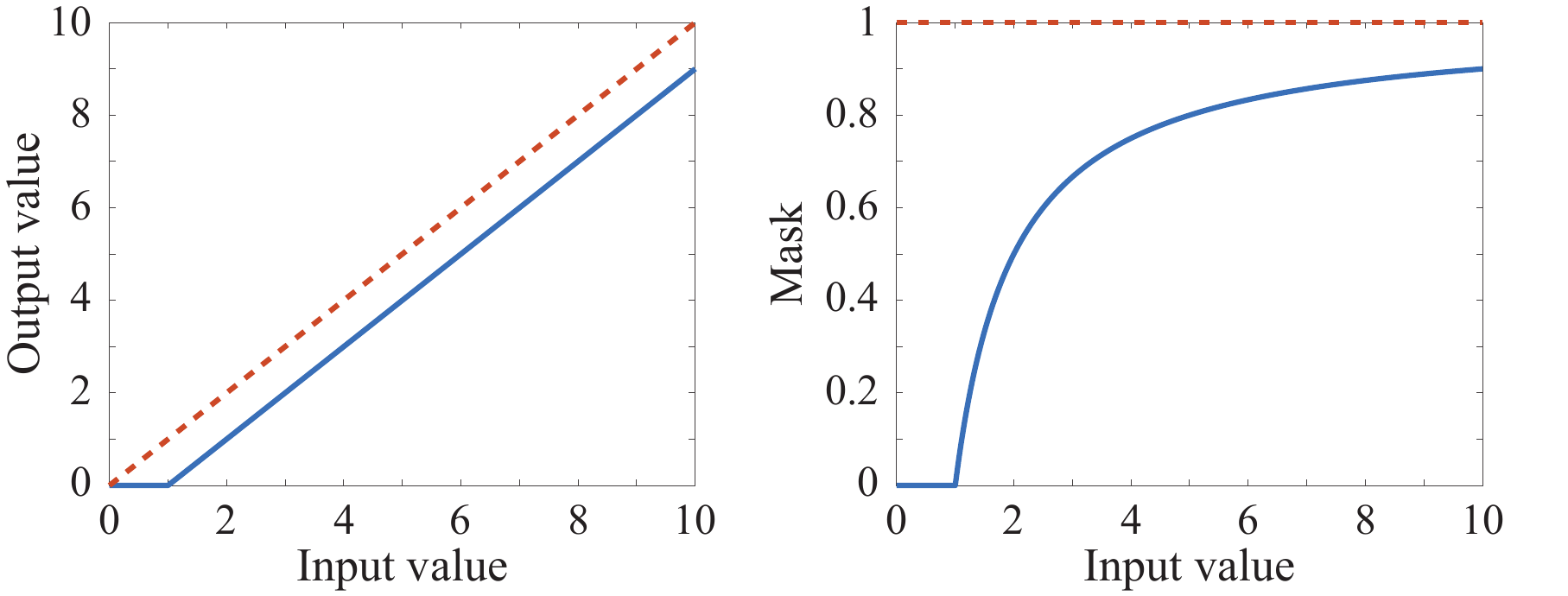}
    \caption{Soft-thresholding and its mask. The soft-thresholding operator shown on the left (solid line) corresponds to the mask in Eq.~\eqref{eq:defL1mask} shown on the right (solid line). The dashed lines represent the identity operator for a reference.}
    \label{fig:softThresholding}
\end{figure}

\subsection{Determined BSS based on Time-frequency Masking}
\label{subsec:defTFmaskingBSS}

The soft- and group-thresholding operators [in Eqs.~\eqref{eq:L1prox} and \eqref{eq:L21prox}] and the generalized thresholding/shrinkage operators [in Eqs.~\eqref{eq:p-shrink} and \eqref{eq:socialShrink}] can be summarized as follows,
\begin{equation}
    \bigl(\mathcal{T}^{\lambda}[\mathbf{z}]\bigr)_n[t,f] = \bigl(\mathcal{M}(\mathbf{z})\bigr)_n[t,f]\; z_{n}[t,f],
\end{equation}
where $0\leq\left(\mathcal{M}(\mathbf{z})\right)_n[t,f]\leq1$ is a non-negative scalar that depends on the input, $\mathbf{z}$.
This process can be interpreted as time-frequency masking using a data-dependent mask $\mathcal{M}(\mathbf{z})$.
For instance, the soft-thresholding operator in Eq.~\eqref{eq:L1prox} is the time-freuqency masking using the following mask:
\begin{equation}
    \bigl(\mathcal{M}_{\ell_1}^{\lambda}(\mathbf{z})\bigr)_n[t,f] = \biggl(1-\frac{\lambda}{|z_n[t,f]|}\biggr)_{\!\!+},
    \label{eq:defL1mask}
\end{equation}
which is illustrated on the right side of Fig.\:\ref{fig:softThresholding}.
Similarly, the group-thresholding operator in Eq.~\eqref{eq:L21prox} uses
\begin{equation}
    \bigl(\mathcal{M}_{\ell_{2,1}}^{\lambda}(\mathbf{z})\bigr)_n[t,f] =\biggl(1-\frac{\lambda}{(\sum_{f=1}^F|z_n[t,f]|^2)^{\!\frac{1}{2}}\!}\biggr)_{\!\!+}
    \label{eq:maskIVA}
\end{equation}
as the mask.
This observation leads us to an idea of substituting a general time-frequency mask into the PDS algorithm so that the mask defines an implicit source model.

\begin{algorithm}[t]
\caption{PDS-BSS-masking \cite{YatabeICASSP2019}}
\label{alg:PDS-BSS-masking}
\begin{algorithmic}[1]
\STATE \textbf{Input:} $\mathbf{X}$, $\mathbf{w}^{[1]}$, $\mathbf{y}^{[1]}$, $\mu_1$, $\mu_2$, $\alpha$
\STATE \textbf{Output:} $\mathbf{w}^{[K+1]}$
\FOR{$k = 1, \ldots, K$}
\STATE $\widetilde{\mathbf{w}} = \prox_{\mu_1 \mathcal{I}}[\;\mathbf{w}^{[k]}-\mu_1\mu_2 \mathbf{X}^\mathrm{H}\mathbf{y}^{[k]}\;]$
\STATE $\;\mathbf{z} = \mathbf{y}^{[k]} + \mathbf{X}(2\widetilde{\mathbf{w}}-\mathbf{w}^{[k]})$
\STATE $\,\widetilde{\mathbf{y}} = \,\mathbf{z} - \mathcal{M}^\theta(\mathbf{z})\odot\mathbf{z}$
\STATE $\,\mathbf{y}^{[k+1]} = \alpha\widetilde{\mathbf{y}}+(1-\alpha)\mathbf{y}^{[k]}$
\STATE $\mathbf{w}^{[k+1]} = \alpha\widetilde{\mathbf{w}}+(1-\alpha)\mathbf{w}^{[k]}$
\ENDFOR
\end{algorithmic}
\end{algorithm}

By heuristically substituting a time-frequency mask for the proximity operator, Algorithm~\ref{alg:PDS-BSS} is extended as Algorithm~\ref{alg:PDS-BSS-masking}, where $\odot$ denotes the element-wise product, and $\theta$ represents a set of parameters for generating the mask.
Although stability and convergence of the algorithm with a general mask can be investigated only by experiments, testing several algorithms is easy because the only effort for rewriting the code is in the 6th line.
One can just insert a masking method into the algorithm and run it for checking the performance.
Any thresholding function and/or sound enhancement method based on time-frequency masking can collaborate with determined BSS through this algorithm, and hence tremendous flexibility is brought by Algorithm~\ref{alg:PDS-BSS-masking}.

This heuristic generalization is closely related to the plug-and-play scheme \cite{Venkatakrishnan2013plugPlay,Chan2017,Ono2017}.
By regarding the definition of the proximity operator in Eq.~\eqref{eq:defProx} as the negative log-likelihood, it can be viewed as a maximum \textit{a posteriori} (MAP) estimator with a prior distribution $C\exp(-\mathcal{P}(\cdot))$ as follows \cite{Venkatakrishnan2013plugPlay}:
\begin{equation}
    \prox_{\lambda\mathcal{P}}[\mathbf{z}] = \arg\max_{\boldsymbol{\xi}}\Bigl[\,\mathrm{e}^{-\frac{1}{2\lambda}\left\|\mathbf{z}-\boldsymbol{\xi}\right\|_2^2} \, \mathrm{e}^{-\mathcal{P}(\boldsymbol{\xi})}\,\Bigr].
    \label{eq:MAP}
\end{equation}
This interpretation suggests that substituting a general Gaussian denoiser, that approximately solves Eq.~\eqref{eq:MAP}, in place of the proximity operator results in an algorithm that works as if the (implicit) function $\mathcal{P}$ is minimized \cite{Venkatakrishnan2013plugPlay,Chan2017,Ono2017}.
When the underlying penalty function is a sum of the penalty functions corresponding to each source $(\mathcal{P} = \sum_{n=1}^N\mathcal{P}_n)$, the algorithm can be interpreted as an independence-based BSS method (ML estimation) with $C\exp(-\mathcal{P}_n(\cdot))$ being the density function of the $n$th source signal \cite{hyvarinen2000}.
In this sense, Algorithm~\ref{alg:PDS-BSS-masking} recasts the BSS problem into the denoising problem in Eq.~\eqref{eq:MAP} with the same prior distribution of the sources.

Note that a BSS algorithm beyond the independence-based framework can also be realized with Algorithm~\ref{alg:PDS-BSS-masking}, at least as a procedure, by inserting a mask that is not separable for each source.
The independence assumption leads to the BSS problem in Eq.~\eqref{eq:logDetProblem} whose source model is given by a sum of the penalty functions.
This is because each source is assumed to be independent from the others, i.e., the source model cannot use information from the other sources to separate a source signal.
Then, the corresponding masking functions are also separately applied to each signal without consideration of the others.
This is a disadvantage of the independence-based framework because full information on all signals cannot be used.
In contrast, Algorithm~\ref{alg:PDS-BSS-masking} allows us to use a mask that simultaneously considers all signals.
We will propose such a non-separable mask in the next section to obtain HVA.

\subsection{Intuitive Interpretation of the Proposed Algorithm}
\label{subsec:intuitiveInterp}

To obtain a better intuition of the working principle, each line of Algorithm~\ref{alg:PDS-BSS-masking} is roughly explained.
For convenience of explanation, we start it from the 6th line.

The 6th line calculates difference between the auxiliary variable, $\mathbf{z}$, and its masked version, $\mathcal{M}^\theta(\mathbf{z})\odot\mathbf{z}$.
Note that the time-frequency masking is applied to spectrograms, and hence $\mathbf{z}$ is some sort of (vectorized) spectrograms.
The difference, $\widetilde{\mathbf{y}}$, contains information about how the masking changed the variable, $\mathbf{z}$.
This information is carried to the 4th line via the 7th and 8th lines.
Since the 7th and 8th lines are weighted averages with variables in the previous iterate, they are merely controllers of speed of the update.
Therefore, the essential information on the difference, $\widetilde{\mathbf{y}}$, is not changed by the 7th line and is brought to the 4th line.

The 4th line is the composition of three operations: multiplication of $\mathbf{X}^\mathrm{H}$, subtraction from $\mathbf{w}^{[k]}$, and application of $\prox_{\mu_1 \mathcal{I}}$.
The multiplication of $\mathbf{X}^\mathrm{H}$ converts the information on masking contained in $\mathbf{y}^{[k]}$ to the domain of the demixing filters, $\mathbf{w}^{[k]}$.
Then, the subtraction updates the demixing filters based on that information.
Since $\mathbf{y}^{[k]}$ is obtained by the difference between the variables before and after masking, it can be regarded as something similar to gradient that informs the effect of masking.
In this sense, the subtraction, $\mathbf{w}^{[k]}-\mu_1\mu_2 \mathbf{X}^\mathrm{H}\mathbf{y}^{[k]}$, updates the demixing filter, $\mathbf{w}^{[k]}$, like the gradient descent method.
It is also updated by $\prox_{\mu_1 \mathcal{I}}$ to avoid the undesired result, e.g., $\mathbf{w}^{[k]}=\mathbf{0}$.

The information on updated  demixing filter, $\widetilde{\mathbf{w}}$, is reflected in the auxiliary variable, $\mathbf{z}$, by the 5th line.
For convenience of explanation, this line is rewritten as $\mathbf{z} = \mathbf{X}\widetilde{\mathbf{w}} + [\mathbf{X}\widetilde{\mathbf{w}} - (\mathbf{X}\mathbf{w}^{[k]} - \mathbf{y}^{[k]})]$.
The first term, $\mathbf{X}\widetilde{\mathbf{w}}$, is the result of filtering applied to the observed data, $\mathbf{X}$, i.e., separated sources.
This filtered signal is modified by the addition of the later terms.
The subtraction of the last two terms, $\mathbf{X}\mathbf{w}^{[k]} - \mathbf{y}^{[k]}$, is similar to that in the 4th line, $\mathbf{w}^{[k]}-\mu_1\mu_2 \mathbf{X}^\mathrm{H}\mathbf{y}^{[k]}$, but is performed in the domain of the auxiliary variable, $\mathbf{y}$. 
By subtracting it from $\mathbf{X}\widetilde{\mathbf{w}}$, the effect of the domain difference and $\prox_{\mu_1 \mathcal{I}}$ is obtained as $[\mathbf{X}\widetilde{\mathbf{w}} - (\mathbf{X}\mathbf{w}^{[k]} - \mathbf{y}^{[k]})]$.
This effect is added to $\mathbf{X}\widetilde{\mathbf{w}}$ and handled by the auxiliary variable, $\mathbf{z}$.
Therefore, the input of the masking, $\mathbf{z}$, consists of not only the separated result for that iteration, $\mathbf{X}\widetilde{\mathbf{w}}$, but also the mismatch between the domains.
This involved structure allows us to consider the masking separately from the demixing filter update.

\subsection{Relation to the Model-based IVA}
\label{sec:modelBasedIVA}

Here, relation between Algorithm~\ref{alg:PDS-BSS-masking} and the model-based IVA \cite{modelBasedIVA} is discussed.
The model-based IVA is an extension of IVA that utilizes a single-channel enhancement method to define an implicit source model.
By considering the time-frequency-variant Gaussian distribution as the source model, with variance $v_n[t,f]$, the penalty function corresponding to the model-based IVA can be written as a weighted $\ell_2$ norm,
\begin{equation}
    \mathcal{P}_n(\mathbf{y}[t,f]) = C\left\|y_n[\cdot,\cdot]\right\|_{2,\mathbf{v}}^2 = C\sum_{t=1}^T\sum_{f=1}^F\frac{|y_n[t,f]|^2}{v_n[t,f]},
\end{equation}
which penalizes a time-frequency bin with small $v_n[t,f]$ more than that with large $v_n[t,f]$.
This variance is chosen as $v_n[t,f] = |\hat{x}_n[t,f]|^2$, where $\hat{x}_n[t,f]$ is a (roughly) separated signal estimated by some single-channel source enhancement method (e.g., spectral subtraction applied to a mixture signal observed by one of the channels \cite{modelBasedIVA}).

As its proximity operator is the shrinkage operator \cite{Patrikh2014},
\begin{equation}
    \bigl(\prox_{\frac{\lambda}{2} \left\|\cdot\right\|_{2,\mathbf{v}}^2}[\mathbf{z}]\bigr)_n[t,f] = \biggl(\frac{v_n[t,f]}{v_n[t,f]+\lambda}\biggr) \, z_n[t,f],
    \label{eq:L2prox}
\end{equation}
the model-based IVA can also be handled by Algorithm~\ref{alg:PDS-BSS-masking} via the mask, $(\mathcal{M}(\mathbf{z}))_n[t,f] = v_n[t,f]/(v_n[t,f]+\lambda)$, which is independent of the inputted variable $\mathbf{z}$ (i.e., constant for every iteration).
Although the two methods are related in terms of using a general time-frequency masking method for estimating the demixing matrix, the model-based IVA utilizes the mask only once, before starting iteration, to calculate the weight, $v_n[t,f] = |(\mathcal{M}(\mathbf{x}))_n[t,f]\,x_n[t,f]|^2$.
In contrast, the proposed algorithm uses the mask within the iteration by updating it based on the inputted variable at that time.
Therefore, the model-based IVA can be regarded as a special case of the proposed masking-based BSS framework.

\begin{figure*}[!t]
    \centering
    \includegraphics[width=1.98\columnwidth]{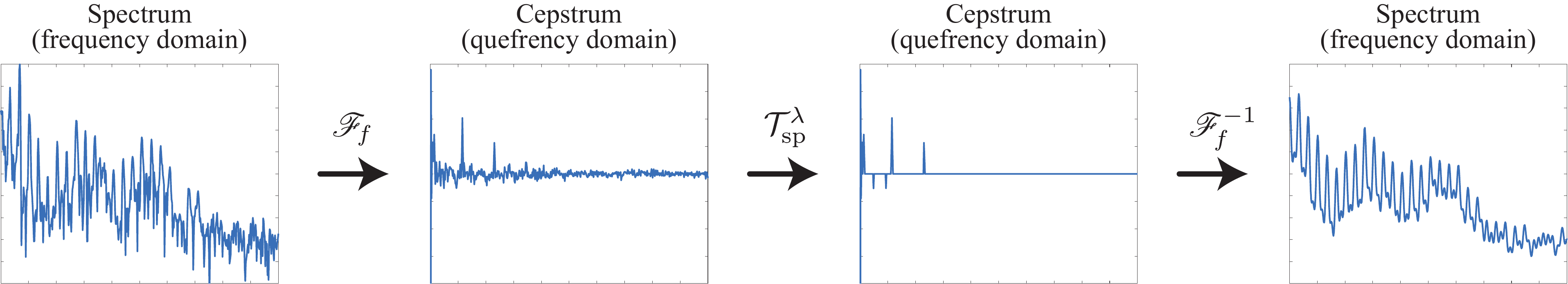}
    \caption{Typical example of a voiced speech signal and its enhancement by the cepstrum (hard-)thresholding. The log-amplitude spectrum of a segment of voiced speech (leftmost) is converted to cepstrum (second from the left) by the Fourier transform. By thresholding the cepstrum coefficients (second from the right) and taking the inverse Fourier transform, the enhanced version of the log-amplitude spectrum is obtained (rightmost) as described in Section \ref{sec:cepThresh}.}
    \label{fig:SchematicCepThresh}
\end{figure*}

\section{Harmonic Vector Analysis (HVA): A Novel BSS Method Based on the Harmonic Structure}
\label{sec:HVA}

As an application of Algorithm~\ref{alg:PDS-BSS-masking}, a BSS method named HVA is proposed in this section.
It is based on some ideas novel for determined BSS, including cepstrum thresholding, non-separable masking, and cosine shrinkage operator.

\subsection{Harmonic Structure of Audio Signals}
\label{sec:harmonicStructure}

In HVA, the harmonic structure is considered as the basis of the mask generation.
As an illustrative example of the harmonic structure, a log-amplitude spectrum of a voiced segment of a speech signal is shown on the left side of Fig.\:\ref{fig:SchematicCepThresh}.
The periodic repetition of the peaks and dips is called harmonic structure and is typical of real-world audio signals.
That is, in a short-time segment of a typical source signal, multiple peaks (or harmonic components) simultaneously occur.
This co-occurrence of the harmonic components should be useful for resolving the permutation problem and separating the source signals because a prominent peak can inform the positions of the other peaks.

To incorporate this prior knowledge into determined BSS based on Algorithm~\ref{alg:PDS-BSS-masking}, a time-frequency masking method should be designed so that the harmonic components are enhanced compared to the other components.
In HVA, this is realized by two ideas that are new to determined BSS: cepstrum thresholding and a Wiener-like mask.

\subsection{Cepstrum Thresholding Enhancing the Harmonic Structure}
\label{sec:cepThresh}

One of the well-accepted concepts related to the harmonic structure is \textit{cepstrum}.
When a log-amplitude spectrum exhibits the harmonic structure, it can be well-approximated by a few Fourier-series coefficients because of the periodic repetition.
To capture such property, cepstrum is defined as the Fourier transform of a log-amplitude spectrum.%
\footnote{In the literature, cepstrum may be defined by the \textit{inverse} Fourier transform of a log-amplitude spectrum. Such difference is not important for HVA since the cepstrum thresholding does not depend on the phase difference.}
By denoting the element-wise absolute value as $(\mathrm{abs}(\mathbf{z}))_n[t,f] = |z_n[t,f]|$, cepstrum of a multi-channel signal, $\mathrm{cep}(\mathbf{z})$, for all time segments and channels can be written as
\begin{equation}
    (\mathrm{cep}(\mathbf{z}))_n[t,c]=(\mathscr{F}_{\!f}(\log(\mathrm{abs}(\mathbf{z}))))_n[t,c],
\end{equation}
where $\log(\cdot)$ is the element-wise logarithmic function, $\mathscr{F}_{\!f}$ is the (normalized) frequency-directional Fourier transform,
\begin{equation}
    (\mathscr{F}_{\!f}(\mathbf{z}))_n[t,c] = \frac{1}{F}\sum_{f=1}^F z_n[t,f] \, \mathrm{e}^{-2\pi\mathbbm{i}(c-1)(f-1)/C},
\end{equation}
and $c=1,\ldots,C$ is the index of quefrency.
Note that zero-padding can be used to make $C>F$, which might improve the performance because of higher redundancy.

By introducing a Fourier thresholding operator $\mathcal{T}_\mathscr{F}^\lambda$ as
\begin{equation}
    \mathcal{T}_\mathscr{F}^\lambda(\mathbf{z}) = \mathscr{F}_{\!f}^{-1}(\mathcal{T}_{\text{sp}}^\lambda(\mathscr{F}_{\!f}(\mathbf{z}))),
    \label{eq:defFthresh}
\end{equation}
we define the \textit{cepstrum thresholding} that applies a sparsity-inducing operator in the cepstrum domain:
\begin{equation}
    \mathcal{T}_\text{cep}^\lambda(\mathbf{z}) = \exp(\mathcal{T}_\mathscr{F}^\lambda(\log(\mathrm{abs}(\mathbf{z})))),
    \label{eq:defCepThresh}
\end{equation}
where $\exp(\cdot)$ is the element-wise exponential function, $\mathscr{F}_{\!f}^{-1}$ is the frequency-directional inverse Fourier transform,
\begin{equation}
    (\mathscr{F}_{\!f}^{-1}(\mathbf{z}))_n[t,f] = \frac{F}{C}\sum_{c=1}^C z_n[t,c] \, \mathrm{e}^{2\pi\mathbbm{i}(c-1)(f-1)/C},
\end{equation}
and $\mathcal{T}_{\text{sp}}^\lambda$ is a sparsity-promoting thresholding operator, e.g., the soft-thresholding operator in Eq.~\eqref{eq:L1prox} or the $p$-shrinkage operator in Eq.~\eqref{eq:p-shrink}.

The effect of the cepstrum thresholding is shown on the right side of Fig.\:\ref{fig:SchematicCepThresh}, where the hard-thresholding was chosen as the cepstrum-domain thresholding operator, $\mathcal{T}_{\text{sp}}^\lambda$, for clear demonstration.
The cepstrum of the log-amplitude spectrum contains some larger peaks that correspond to the harmonic components.
The other small cepstrum coefficients correspond to the non-structured details.
By removing small cepstrum coefficients and retaining larger coefficients using hard-thresholding, the harmonic structure is enhanced as in the rightmost figure.
For mixture signals, the cepstrum thresholding enhances the dominant signal having harmonic structure and attenuates the other components.
We expect that, for each channel, such attenuation principally occurs to the interference signals that have less energy than the main signal.

\subsection{Non-separable Masking for Source Separation}
\label{sec:multiMask}

Based on a set of enhanced signals, a mask is constructed and applied in Algorithm~\ref{alg:PDS-BSS-masking}.
Ordinarily, the independence criterion has forced a BSS algorithm to be a procedure separable for each source signal.
In contrast, it is also possible to define a non-separable BSS method that simultaneously considers all source signals to extract separation cues.
Algorithm~\ref{alg:PDS-BSS-masking} can realize such method by using a non-separable mask-generating function, for example, the Wiener-like mask \cite{ErdoganICASSP2015}:
\begin{equation}
    (\mathcal{M}_{\text{WL}}(\hat{\mathbf{x}}))_n[t,f] = \biggl(\frac{|\hat{x}_n[t,f]|^2}{\sum_{n=1}^N|\hat{x}_n[t,f]|^2}\biggr)^{\!\!\gamma},
    \label{eq:defNonSepMask}
\end{equation}
which takes values between 0 and 1, where $\hat{x}_n[\cdot,\cdot]$ is the enhanced spectrogram corresponding to the $n$th source signal, and $\gamma>0$ is a parameter adjusting the level of attenuation.
Note that, when $\gamma=1$, this mask can be viewed as a non-separable version of the mask in Eq.~\eqref{eq:L2prox}, which is related to the time-frequency-variant Gaussian model, by replacing the constant $\lambda$ in the denominator of Eq.~\eqref{eq:L2prox} with the sum of the other source signals, $\sum_{l\neq n}|\hat{x}_l[t,f]|^2$.

This non-separable mask is more effective for promoting source separation than ordinary separable masks because it simultaneously uses information on all signals and encourages each bin to be more exclusive and unmixed.

\subsection{Harmonic Vector Analysis (HVA)}
\label{sec:HVAexplain}

We propose HVA by defining a specific mask that uses the harmonic structure for enhancing the source signals.
It is a combination of the cepstrum thresholding and Wiener-like masking introduced in the previous subsections:
\begin{equation}
    (\mathcal{M}_{\text{HVA}}^{\lambda,\kappa}(\mathbf{z}))_n[t,f] = \biggl(\frac{\upsilon_n^{\mathbf{z},\lambda,\kappa,\varepsilon}[t,f]}{\sum_{n=1}^N\upsilon_n^{\mathbf{z},\lambda,\kappa,\varepsilon}[t,f]}\biggr)^{\!\!\gamma},
    \label{eq:defHVAmaskFun}
\end{equation}
where $\upsilon_n^{\mathbf{z},\lambda,\kappa,\varepsilon}[\cdot,\cdot]$ is a squared amplitude spectrogram whose harmonic structure is enhanced by the cepstrum thresholding [corresponding to $|\hat{x}_n[\cdot,\cdot]|^2$ in Eq.~\eqref{eq:defNonSepMask}].

The cepstrum thresholding in Eq.~\eqref{eq:defCepThresh} is composition of $\exp$, $\mathcal{T}^{\lambda}_\mathscr{F}$, $\log$ and $\mathrm{abs}$.
Therefore, $\upsilon_n^{\mathbf{z},\lambda,\kappa,\varepsilon}[\cdot,\cdot]$ in Eq.~\eqref{eq:defHVAmaskFun} is explained in this order.
By applying the exponential function to a thresholded log-amplitude spectrogram, $\varrho_n^{\mathbf{z},\lambda,\kappa,\varepsilon}[\cdot,\cdot]$, and squaring it, $\upsilon_n^{\mathbf{z},\lambda,\kappa,\varepsilon}[\cdot,\cdot]$ is obtained as follows:
\begin{equation}
    \upsilon_n^{\mathbf{z},\lambda,\kappa,\varepsilon}[t,f] = \exp(2\varrho_n^{\mathbf{z},\lambda,\kappa,\varepsilon}[t,f]),
    \label{eq:HVAlastStepExp}
\end{equation}
where $2$ comes from the squaring in Eq.~\eqref{eq:defNonSepMask}.
The thresholded log-amplitude spectrogram, $\varrho_n^{\mathbf{z},\lambda,\kappa,\varepsilon}[\cdot,\cdot]$, is given by Fourier thresholding, $\mathcal{T}_\mathscr{F}^{\lambda,\kappa}(\cdot)=\mathscr{F}_{\!f}^{-1}(\mathcal{T}_{\text{sp}}^{\lambda,\kappa}(\mathscr{F}_{\!f}(\cdot)))$, applied to mean-subtracted log-amplitude spectrograms, $\boldsymbol{\rho}^{\mathbf{z},\varepsilon}$, as follows:
\begin{equation}
    \varrho_n^{\mathbf{z},\lambda,\kappa,\varepsilon}[t,f] = (\mathcal{T}^{\lambda,\kappa}_\mathscr{F}(\boldsymbol{\rho}^{\mathbf{z},\varepsilon}))_n[t,f]+\mu_n^{\mathbf{z},\varepsilon}[t],
    \label{eq:HVAmeanAdd}
\end{equation}
where $\boldsymbol{\rho}^{\mathbf{z},\varepsilon}$ is obtained by subtracting time-dependent mean,
\begin{equation}
    \rho_n^{\mathbf{z},\varepsilon}[t,f] = \log(|z_n[t,f]|+\varepsilon) - \mu_n^{\mathbf{z},\varepsilon}[t],
    \label{eq:HVAmeanSubs}
\end{equation}
$\mu_n^{\mathbf{z},\varepsilon}[t]$ is the mean value of the log-amplitude spectrum,
\begin{equation}
    \mu_n^{\mathbf{z},\varepsilon}[t] = \frac{1}{F}\sum_{f=1}^F\log(|z_n[t,f]|+\varepsilon),
\end{equation}
and $\varepsilon>0$ is a small constant for preventing $\rho_n^{\mathbf{z},\varepsilon}[t,f]$ to be $-\infty$.
Mean subtraction is performed in Eq.~\eqref{eq:HVAmeanSubs} to make the cepstrum coefficients sparser.
It also has another benefit that adding the mean value after the thresholding as in Eq.~\eqref{eq:HVAmeanAdd} can restore the level of log-amplitude.
Therefore, the Wiener-like mask can enhance the level difference of the signals even when $\lambda$ is exceedingly large and the thresholder eliminates all cepstrum coefficients.

\begin{figure}[!t]
    \centering
    \includegraphics[width=0.99\columnwidth]{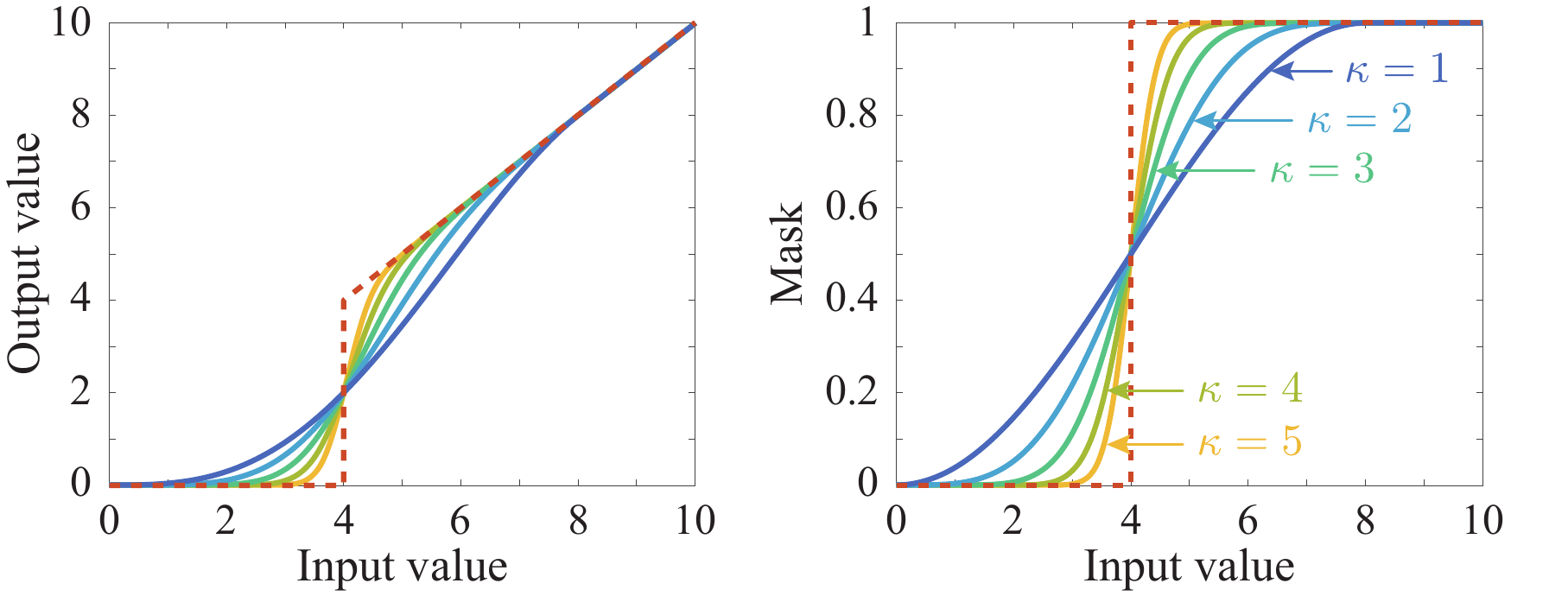}
    \caption{Cosine shrinkage and its mask for $\lambda=4$ and $\kappa\in\{1,2,3,4,5\}$. The cosine shrinkage operator in Eq.~\eqref{eq:defCosThresh} (solid line) is shown on the left with the hard-thresholding operator (dashed line). The corresponding masks (the composition of raised cosine functions drawn by solid lines, and the step function drawn by dashed line) are shown on the right.}
    \label{fig:cosineThresholding}
\end{figure}

For the cepstrum-domain operator, $\mathcal{T}_{\text{sp}}^{\lambda,\kappa}$, included in $\mathcal{T}_\mathscr{F}^{\lambda,\kappa}$, any sparsity-promoting operator can be adopted.
In this paper, we newly propose the following operator, which is named \textit{cosine shrinkage operator}, for the enhancement:
\begin{equation}
    (\mathcal{T}^{\lambda,\kappa}_{\text{cos}}(\mathbf{z}))_n[t,c] = 
    \Xi^\kappa[(|z_n[t,c]|/(2\lambda))_{\overline{1}}]\:z_n[t,c],
    \label{eq:defCosThresh}
\end{equation}
where $\Xi^\kappa$ is $\kappa$-times composition of $\Xi$ (i.e., $\Xi^\kappa=\Xi\circ\cdots\circ\Xi$),
\begin{equation}
    \Xi\,[z_n[t,c]] = (1-\cos(\pi z_n[t,c]))/2,
\end{equation}
and $(\cdot)_{\overline{1}}=\min\{1,\cdot\}$.
For intuitive explanation, it is illustrated in Fig.~\ref{fig:cosineThresholding} with the corresponding mask, $\Xi^\kappa[(|z_n[t,c]|/(2\lambda))_{\overline{1}}]$.
The mask (on the right side of Fig.~\ref{fig:cosineThresholding}) consists of the half period of (raised) cosine function, where $\lambda$ is its inflection point.
This can be viewed as a smooth approximation of the mask corresponding to the hard-thresholding operator (red dashed line), and the degree of approximation is controlled by $\kappa$.
Hence, $\mathcal{T}^{\lambda,\kappa}_{\text{cos}}$ is a smooth approximation of the hard-thresholding operator as on the left side of Fig.~\ref{fig:cosineThresholding}.
The reasons why this shrinkage operator is adopted in HVA are as follows: (1) it has no bias for large coefficients similar to hard-thresholding; (2) we found that smoothness is important for stable separation; and (3) HVA does not require to force small coefficients to be exactly zero owing to the Wiener-like mask.
Note again that any thresholding-like function can be used in place of this shrinkage operator.
The performance of HVA depends on its choice, and some other thresholding/shrinkage operator that performs better for HVA than the cosine shrinkage operator should exist.

The proposed algorithm for HVA is shown in Algorithm~\ref{alg:HVA}.%
\footnote{Our MATLAB implementation is available at \cite{HVAcode}.}
For assisting implementation, the mask-generating function, $\mathcal{M}_{\text{HVA}}^{\lambda,\kappa}(\cdot)$ in the $6$th line, is also summarized in Algorithm~\ref{alg:HVAmask}.
The subtraction and addition of the mean value, $\mu_n^{\mathbf{z},\varepsilon}[t]$, in the 5th and 12th lines, respectively, maintain the energy of the squared amplitude spectrograms, $\upsilon_n^{\mathbf{z},\lambda,\kappa,\varepsilon}[\cdot,\cdot]$, similar to that of the input spectrogram, $|z_n[\cdot,\cdot]|^2$.
The (frequency-directional) Fourier transform, $\mathscr{F}_{\!f}$ in the 6th line, converts the log-amplitude spectrum into cepstrum, and its inverse, $\mathscr{F}_{\!f}^{-1}$ in the 11th line, does the opposite.
The mask of the cosine shrinkage operator (the right figure of Fig.\:\ref{fig:cosineThresholding}) is denoted by $\boldsymbol{\varsigma}$ and is applied to the cepstrum coefficients, $\boldsymbol{\nu}$, as $\boldsymbol{\varsigma}\odot\boldsymbol{\nu}$ in the 11th line.
The exponential in the 13th line cancels the logarithm in the 4th line, and the 14th line computes the Wiener-like mask.
Note that $\varsigma_n[t,c]=1$ when $\lambda=0$.
That is, $\upsilon_n[t,f] = |z_n[t,f]|^2$ when $\lambda=0$ (and $\varepsilon=0$).

\begin{algorithm}[t]
\caption{Harmonic Vector Analysis (HVA)}
\label{alg:HVA}
\begin{algorithmic}[1]
\STATE \textbf{Input:} $\mathbf{X}$, $\mathbf{w}^{[1]}$, $\mathbf{y}^{[1]}$, $\mu_1$, $\mu_2$, $\alpha$
\STATE \textbf{Output:} $\mathbf{w}^{[K+1]}$
\FOR{$k = 1, \ldots, K$}
\STATE $\widetilde{\mathbf{w}} = \prox_{\mu_1 \mathcal{I}}[\;\mathbf{w}^{[k]}-\mu_1\mu_2 \mathbf{X}^\mathrm{H}\mathbf{y}^{[k]}\;]$
\STATE $\,\mathbf{z}\, = \mathbf{y}^{[k]} + \mathbf{X}(2\widetilde{\mathbf{w}}-\mathbf{w}^{[k]})$
\STATE $\:\widetilde{\mathbf{y}} = \,\mathbf{z} - \mathcal{M}_{\text{HVA}}^{\lambda,\kappa}(\mathbf{z})\odot\mathbf{z}$
\STATE $\,\mathbf{y}^{[k+1]} = \alpha\widetilde{\mathbf{y}}+(1-\alpha)\mathbf{y}^{[k]}$
\STATE $\mathbf{w}^{[k+1]} = \alpha\widetilde{\mathbf{w}}+(1-\alpha)\mathbf{w}^{[k]}$
\ENDFOR
\end{algorithmic}
\end{algorithm}

\begin{algorithm}[t]
\caption{Computation of $\mathcal{M}_{\text{HVA}}^{\lambda,\kappa}(\mathbf{z})$}
\label{alg:HVAmask}
\begin{algorithmic}[1]
\STATE \textbf{Input:} $\mathbf{z}$, $\lambda$, $\kappa$
\STATE \textbf{Output:} $\mathcal{M}_{\text{HVA}}^{\lambda,\kappa}(\mathbf{z})$
\STATE $\boldsymbol{\zeta} = \log(\mathrm{abs}(\mathbf{z})+\varepsilon)$
\STATE $\mu_n[t] = (1/F)\sum_{f=1}^F\zeta_n[t,f]\quad\forall n,t$
\STATE $\rho_n[t,f] = \zeta_n[t,f] - \mu_n[t]\quad\forall n,t,f$
\STATE $\boldsymbol{\nu}=\mathscr{F}_{\!f}(\boldsymbol{\rho})$
\STATE $\varsigma_n[t,c] = \min\{1,|\nu_n[t,c]|/(2\lambda)\}\quad\forall n,t,c$
\FOR{$k = 1,\ldots,\kappa$}
\STATE $\varsigma_n[t,c] = (1-\cos(\pi\varsigma_n[t,c]))/2\quad\forall n,t,c$
\ENDFOR
\STATE $\boldsymbol{\xi} = \mathscr{F}_{\!f}^{-1}(\boldsymbol{\varsigma}\odot\boldsymbol{\nu})$
\STATE $\varrho_n[t,f] = \xi_n[t,f]+\mu_n[t]\quad\forall n,t,f$
\STATE $\upsilon_n[t,f] = \exp(2\varrho_n[t,f])\quad\forall n,t,f$
\STATE $(\mathcal{M}_{\text{HVA}}^{\lambda,\kappa}(\mathbf{z}))_n[t,f] = \bigl(\upsilon_n[t,f]/\sum_{n=1}^N\upsilon_n[t,f]\bigr)^{\!\gamma}\;\;\forall n,t,f$
\end{algorithmic}
\end{algorithm}

To reduce the number of parameters, $\varepsilon$ is fixed and omitted from $\mathcal{M}_{\text{HVA}}^{\lambda,\kappa}$ because, according to our preliminary investigation, its effect to the performance is not notable ($\varepsilon$ is set to $10^{-3}$ in the rest of the paper).
We also heuristically fix $\gamma$ to $1/N$ in this paper based on the following reason.
When $\upsilon_n^{\mathbf{z},\lambda,\kappa,\varepsilon}[t,f]$ is the same for all $n$, the value inside the parentheses of Eq.~\eqref{eq:defHVAmaskFun} is $1/N$, which depends on $N$.
By setting $\gamma=1/N$, the value of the mask for that case becomes $(1/N)^{(1/N)}\approx0.7$ that is approximately independent of $N$.
Therefore, to avoid decrease in the average value of the mask, $\gamma=1/N$ is chosen tentatively.%
\footnote{%
We empirically found that the algorithm becomes unstable when the average value of the mask is small.
This should be because the PDS algorithm is built upon the proximity operator that has restriction on the amount of change of the signal.
To reduce the amount of change caused by the masking, the values of the mask should be close to $1$.
Therefore, we chose $(1/N)^{(1/N)}\approx0.7$ that is relatively close to $1$.
Although we chose $(\cdot)^\gamma$ for the definition of the Wiener-like mask in Eq.~\eqref{eq:defNonSepMask} because it seems popular in the literature of time-frequency masking, any function that controls the distribution of the value of the mask can be used in place of $(\cdot)^\gamma$.
Investigation of such function that performs better than $(\cdot)^\gamma$ can be a part of future works.
}
The other two parameters, $\lambda$ and $\kappa$, should be chosen based on the distribution of the cepstrum coefficients of the observed signals, which will be investigated in the experimental section.

\begin{figure}[!t]
    \centering
    \includegraphics[width=0.99\columnwidth]{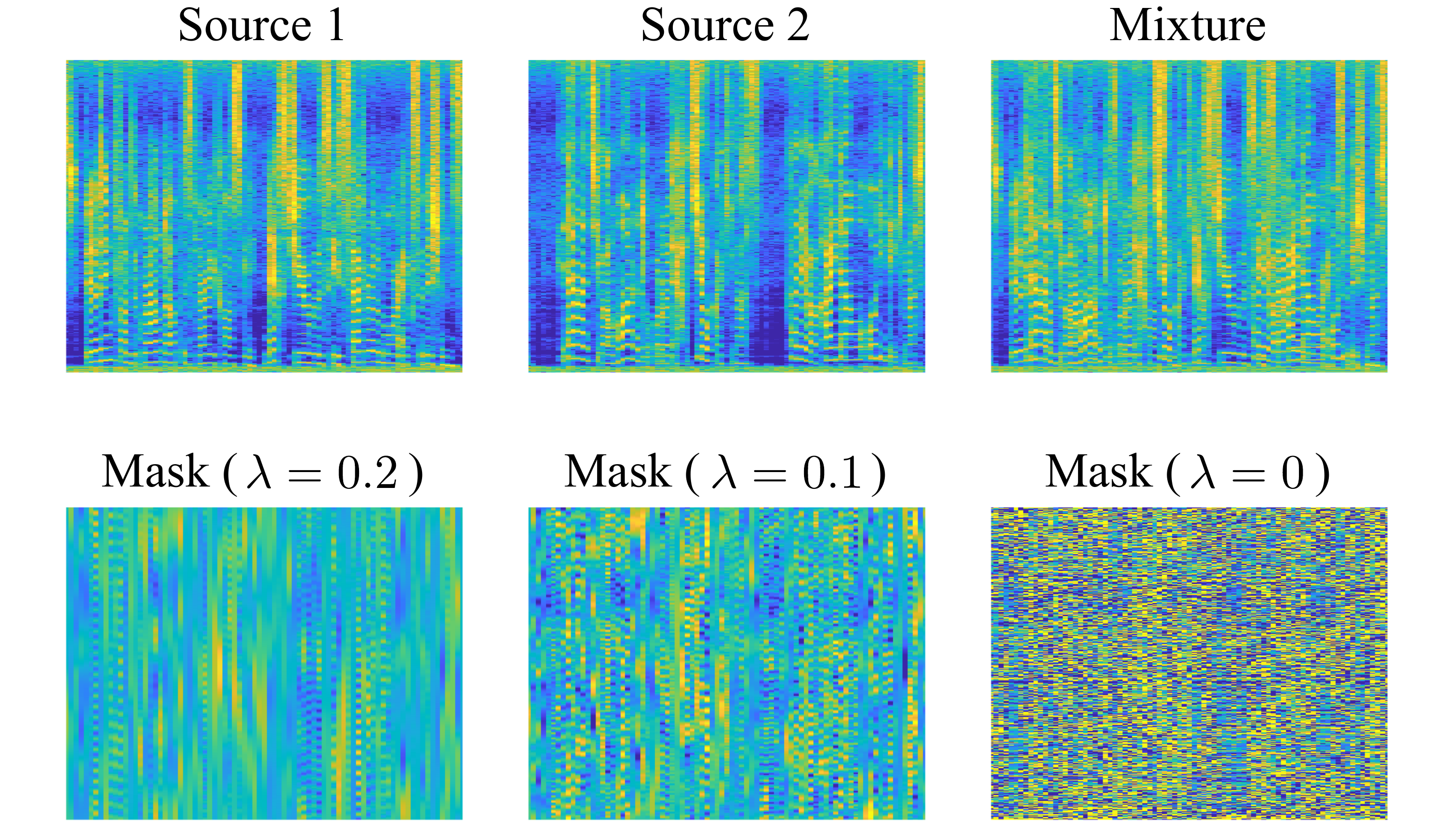}
    \caption{Whitened spectrograms and masks of HVA. The upper row shows a log-amplitude spectrogram of a speech mixture (right) together with its source signals (left and center). The color range is $60$\,dB. Due to the frequency-wise whitening, the signal-to-noise ratio seems different for each frequency. The bottom row shows the masks of HVA (with different $\lambda$s) calculated from the (top right) mixture. The color range is set to $[0.55, 0.85]$ for enhancing the visibility (note that $\gamma=1/N$ makes the average around $0.7$, and therefore $0.7\pm0.15$ was chosen for the range). These masks correspond to the 1st iteration of HVA (see Fig.\:\ref{fig:maskEachIter} for similar masks with color range $[0, 1]$).}
    \label{fig:maskExample}
\end{figure}

\subsection{Role of the Cepstrum Thresholding in HVA}
\label{subsec:roleOfCepThresh}

For demonstration of the mask of HVA, some examples are shown in the bottom row of Fig.~\ref{fig:maskExample}.
Each figure illustrates one of the 2-channel signals/masks, and the corresponding masks for the other channel are not shown here.

The bottom-right figure shows the mask without cepstrum thresholding (i.e., $\lambda=0$).
Note that the non-separable mask itself can promote separation if, for each time-frequency bin, level difference between the channels exists, because the mask retains louder components and attenuates smaller components.
Therefore, the mask in the bottom-right figure is not totally random but exhibits some structure.
However, this mask cannot solve the permutation problem because each frequency is treated independently.

The cepstrum thresholding assists the non-separable mask by enhancing the dominant periodic pattern corresponding to the harmonic structure.
Since the cosine shrinkage operator attenuates cepstrum coefficients that are small relative to the parameter $\lambda$, a larger $\lambda$ gives a simpler mask that can be well-described by fewer sinusoidal patterns.
These examples show that the cepstrum thresholding is not intended to separate some components but just enhancing the harmonic structure.
Note that subtraction of the mean value, $\mu_n^{\mathbf{z},\varepsilon}[t]$, in the logarithmic domain normalizes the time-segment-wise scale of the input spectrogram because $\log(\mathrm{abs}(c\mathbf{z}))=\log(\mathrm{abs}(\mathbf{z})) + \log(c)$ for any positive constant $c>0$.
Therefore, the shrinkage parameter, $\lambda$, can be chosen without a care of the scale of the observed signals.

\begin{figure}[!t]
    \centering
    \includegraphics[width=0.99\columnwidth]{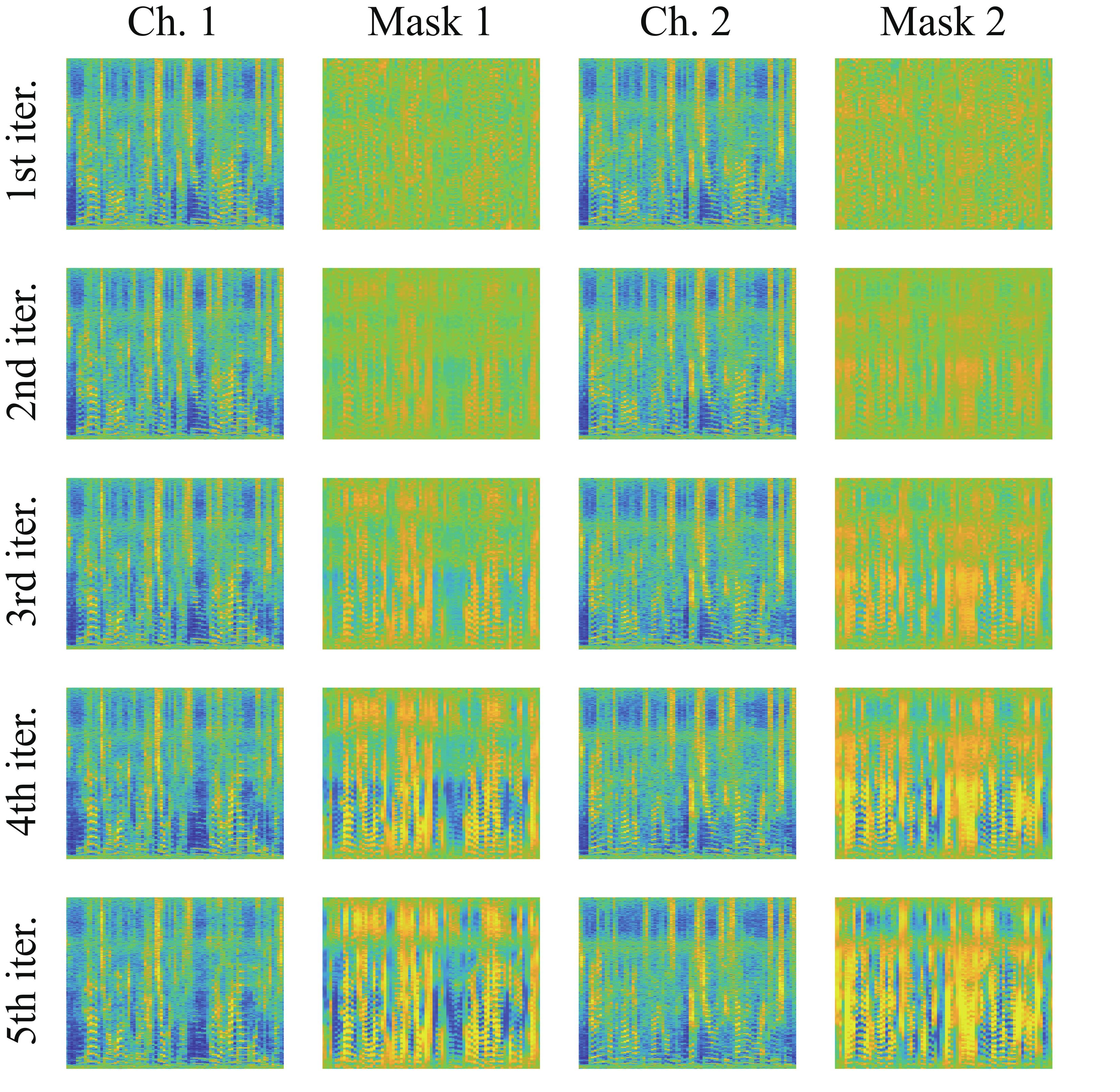}
    \caption{Whitened spectrograms and masks for the first 5 iterations of HVA. The parameters of HVA were set to $\alpha=\mu_1=\mu_2=1$, $\lambda=0.08$, and $\kappa=3$. The signal is Mixture A in Fig.\:\ref{fig:micSpeakerSettingsSpeech}. Therefore, these spectrograms and masks correspond to the top central figures in Figs.\:\ref{fig:IVAparamAlpha} and \ref{fig:IVAparamMu}. The color range for the whitened spectrograms is 60\,dB, and that for the masks is $[0,1]$.}
    \label{fig:maskEachIter}
\end{figure}

For further demonstration, the mask of HVA for each iteration is shown in Fig.\:\ref{fig:maskEachIter}.
This figure is visualization of the first 5 iterations of the experimental result ($\alpha=\mu_1=\mu_2=1$, Mixture A) in Section \ref{subsec:HVAbasicProp}.
At the 1st iteration, the masks seem somewhat random because the cepstrum thresholding was performed for each time segment independently.
Then, the update of the demixing filter collects the information of the mask for all time segments via $\mathbf{X}^\mathrm{H}\mathbf{y}$ (see Algorithm \ref{alg:X'y}).
By using the updated demixing filter, each signal was enhanced, which made the cepstrum thresholding able to adapt to the spectral patterns better in the later iterations.
As can be seen in the figure, the mask rapidly captured the spectral patterns in the first few iterations.

Since the cepstrum thresholding simultaneously processes all frequency components, the mask exhibited vertical patterns.
That is, the effect of the demixing filer at some frequency propagates vertically to all the other frequencies.
Such masks that simultaneously handle all frequencies can solve the permutation problem to some extent.
Note that, since this paper considers the time-invariant model in Eq.~\eqref{eq:mixingProcess}, permutation across the time does not occur.
If the cepstrum thresholding is used in a time-varying situation, the permutation across the time should be treated by some additional technique.

\section{Experiments}
\label{sec:experiment}

In this paper, we presented the general BSS algorithm and its specific application termed HVA.
To show the properties of both the algorithm and HVA, some experiments are conducted in this section.
At first, the properties of the algorithm and HVA are qualitatively shown using two 2-channel speech mixtures as examples.
Then, the performances of HVA over speech and music mixtures in 2- and 3-channel conditions are compared with IVA and ILRMA quantitatively.%
\footnote{An audio example for Section \ref{subsec:speechExp} is available at \cite{KitamuraWeb}.}

The performance was measured by the standard metrics: the source-to-distortion ratio (SDR), source-to-interferences ratio (SIR), and sources-to-artifacts ratio (SAR) \cite{Vincent2006}.
For all trials, the initial value of the demixing matrices $\mathbf{w}^{[1]}$ was set to the identity matrices ($\mathbf{W}[f]=\mathbf{I}$ for all $f$), and that of $\mathbf{y}$ was the zero vector.
The sampling rate of the signals was $16$\,kHz.
The window length was set according to the previous studies for easier comparison.%
\footnote{The window length determines the degree of freedom of the demixing filter. Therefore, the ideal separation performance is higher when the window length is longer. However, a longer window results in more optimization variables, which makes the optimization more difficult. Moreover, the window length determines the appearance of the spectrogram, which makes the characteristics of the source model different. Because of these factors, the relation between the window length and separation performance is very complicated as indicated in \cite{ConsistentILRMA}. In this paper, we decided to follow the previous studies for the window length so that such complication is avoided.}
The whitening \cite{hyvarinen2004independent} and back projection \cite{Matsuoka2002projBack} were applied as pre- and post-processes, respectively.

\subsection{Illustration of Basic Properties of the PDS Algorithm}
\label{subsec:basicProp}

\begin{figure}[!t]
    \centering
    \includegraphics[width=0.92\columnwidth]{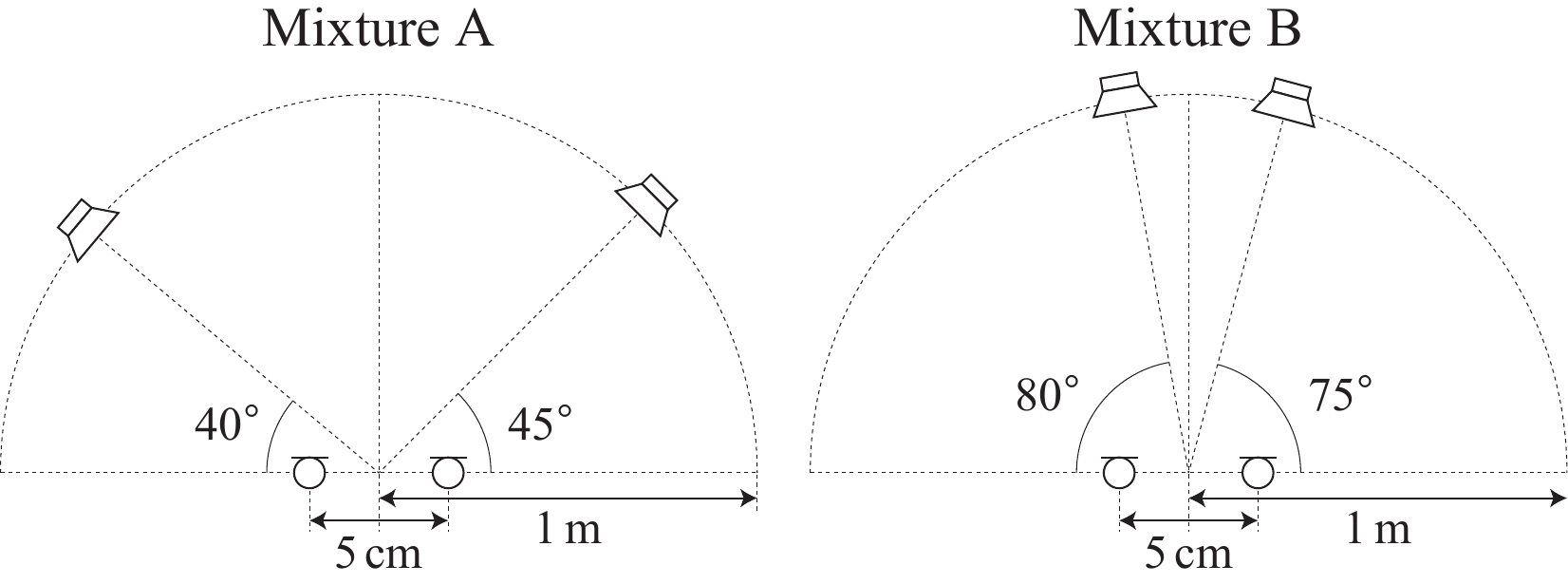}
    \caption{Recording conditions of the 2-channel speech mixtures utilized in the experiments in Sections~\ref{subsec:basicProp} and \ref{subsec:HVAbasicProp} using the Laplace IVA.}
    \label{fig:micSpeakerSettingsSpeech}
\end{figure}

The PDS algorithms contain three parameters $\mu_1$, $\mu_2$, and $\alpha$.
At first, their effects to the performance over iteration are presented using the well-understood IVA.
Two pairs of female speech signals recorded as in Fig.\:\ref{fig:micSpeakerSettingsSpeech} were downloaded from SiSEC database \cite{SAraki2012_SiSEC} (\texttt{liverec} of \texttt{dev1} in the underdetermined audio source separation task),
where the reverberation time was 130\,ms.
The half-overlapping 2048-point-long Hann window (128\,ms) was used for the short-time Fourier transform (STFT).
The BSS method tested here was IVA based on the spherical Laplace distribution, $\mathcal{P}_n=\|\cdot\|_{2,1}$ in Eq.~\eqref{eq:ell21}, whose proximity operator is given in Eq.~\eqref{eq:L21prox}.
Note that the mixture signals contain ambient noise of the room.

\subsubsection{Appropriateness of the PDS algorithm}
\label{subsubsec:compareMMandPDS}

\begin{figure}[!t]
    \centering
    \includegraphics[width=0.9\columnwidth]{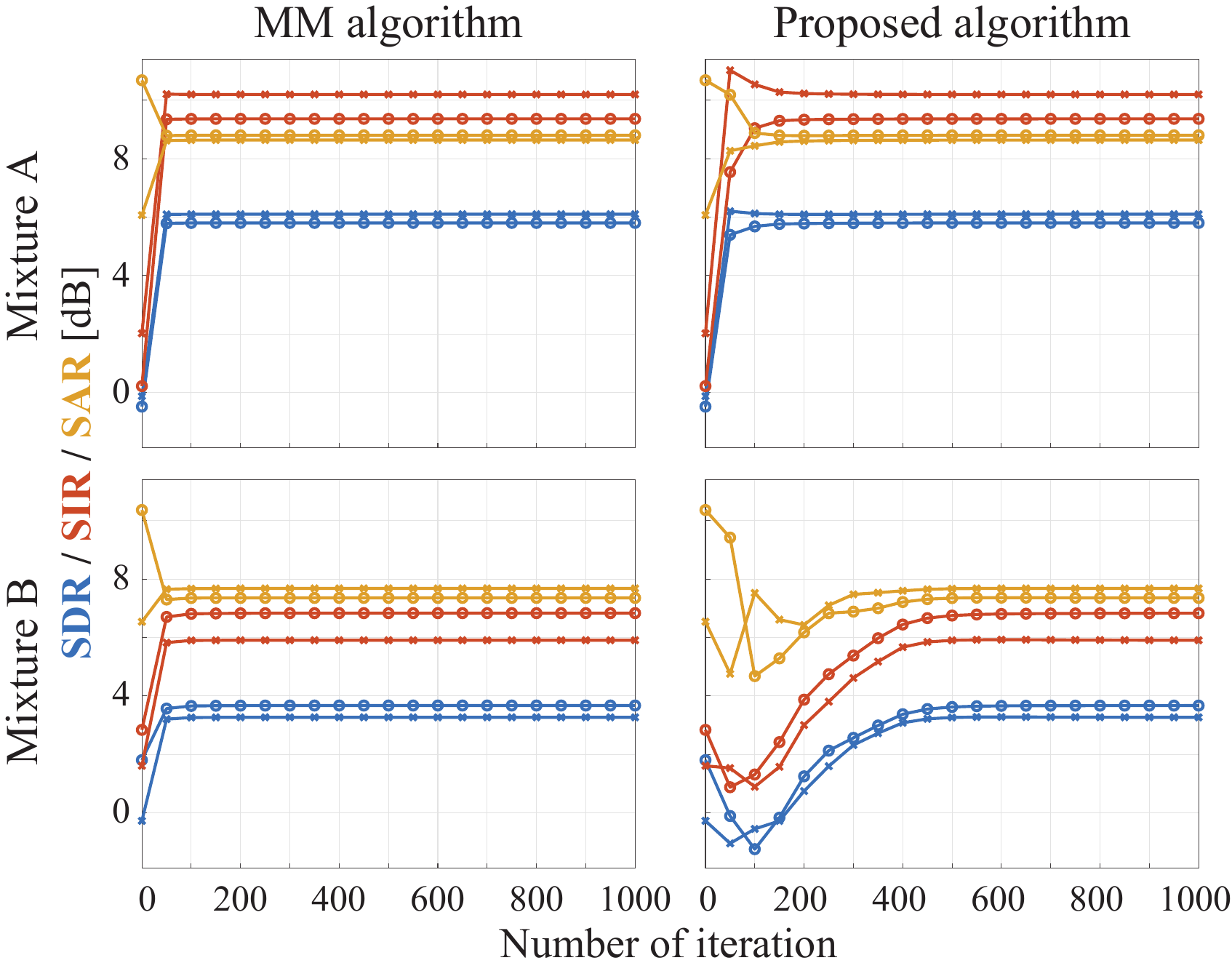}
    \caption{SDR/SIR/SAR of the signals separated from the mixtures (recorded as in Fig.\:\ref{fig:micSpeakerSettingsSpeech}) using Laplace IVA (Section \ref{subsubsec:compareMMandPDS}). The two sources in each mixture are distinguished by the markers (\texttt{o} and \texttt{x}). The left and right figures were obtained by auxIVA \cite{Ono2011auxIVA} and the proposed algorithm (Algorithm~\ref{alg:PDS-BSS}), respectively.
    The parameters were set to $\mu_1=\mu_2=1$ and $\alpha=1.75$.}
    \label{fig:auxIVAvsProxIVA}
\end{figure}

Before showing the effects of parameters, the PDS algorithm was compared with the MM algorithm (auxIVA \cite{Ono2011auxIVA}), based on the iterative projection technique \cite{Ono2010auxFDICA}, to confirm that the Laplace IVA was appropriately realized by the PDS algorithm.
Their performances over iteration are shown in Fig.\:\ref{fig:auxIVAvsProxIVA}, where the parameters were set to $\mu_1=\mu_2=1$, and $\alpha=1.75$.
As in the figure, both algorithms resulted in the same scores, which indicates that the PDS algorithm was properly working, but the PDS algorithm required more iterations than the MM algorithm.
In particular, Mixture B needed significantly more iterations, which should be because it was obtained with a condition more difficult than Mixture A, as in Fig.\:\ref{fig:micSpeakerSettingsSpeech}.
Note that the computation per iteration of IVA by the PDS algorithm was 1.3 times faster than IVA by the MM algorithm (PDS: 26.7 ms, MM: 35.0 ms) which should be because the MM algorithm calculates a lot of matrix inversions within an iteration.

\begin{figure}[!t]
    \centering
    \includegraphics[width=0.98\columnwidth]{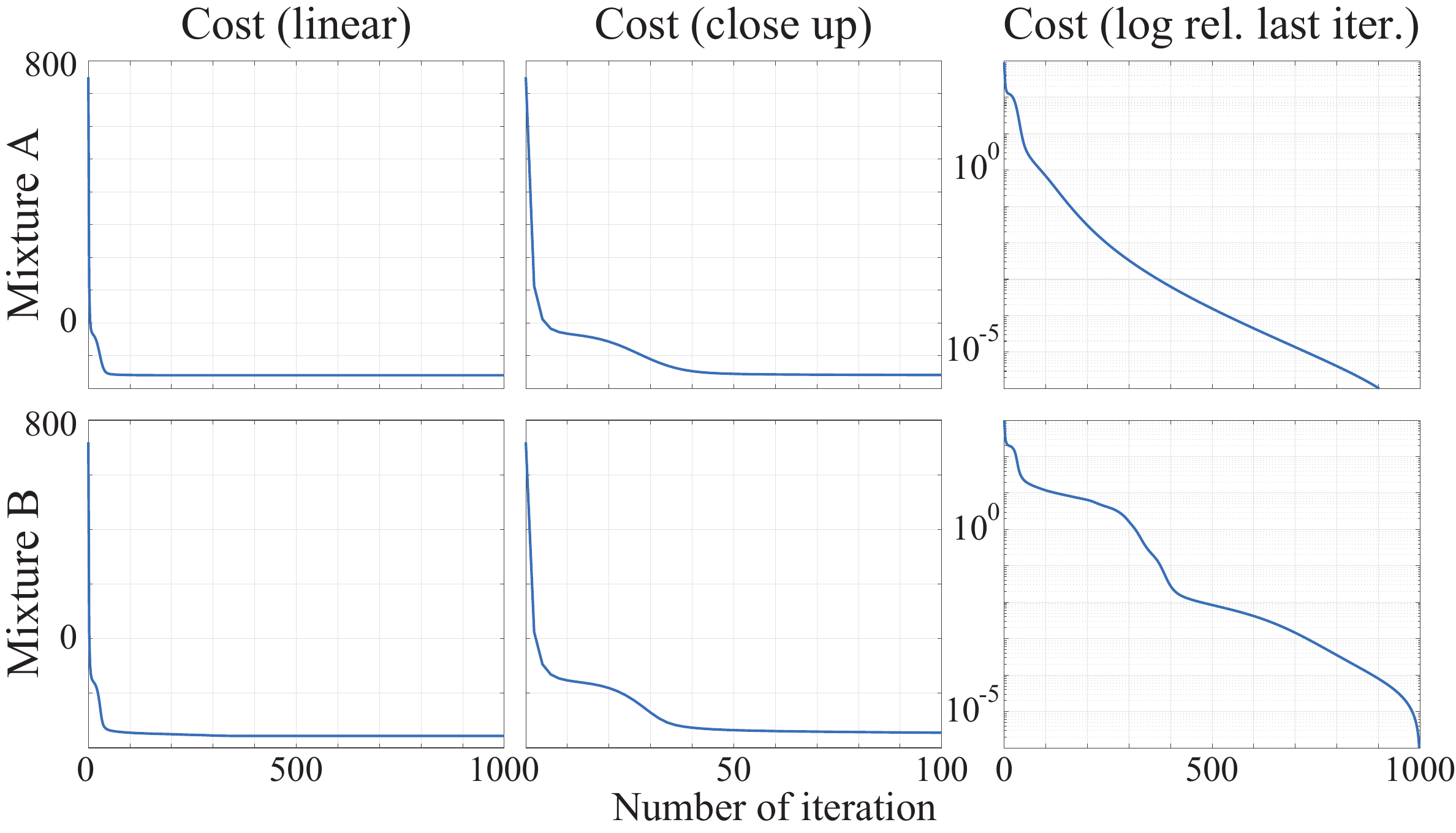}
    \caption{Values of the objective function. With the penalty function in Eq.~\eqref{eq:ell21}, the objective function is given in Eq.~\eqref{eq:logDetProblem}.
    The setting was the same as that of Fig.~\ref{fig:auxIVAvsProxIVA}. The first 100 iterations of the left figure are closed up in the central figure. The right figure shows the same curve in the logarithmic scale with some vertical shift that made the value at the 1000th iteration 0.}
    \label{fig:costProxIVA}
\end{figure}

To see the appropriateness in terms of minimization, the values of the objective function for the PDS algorithm are shown in Fig.~\ref{fig:costProxIVA} with 3 different axes.
They confirm that the proposed algorithm properly reduced the objective function in Eq.~\eqref{eq:logDetProblem} with $\mathcal{P}_n=\|\cdot\|_{2,1}$ in Eq.~\eqref{eq:ell21}.
Note that, in general, this kind of plot cannot be drawn for the proposed masking-based algorithm because an explicit form of the objective function may not exist.

\subsubsection{Effect of the Relaxation Parameter $\alpha$}
\label{subsubsec:effectAlpha}

As mentioned in Section~\ref{sec:PDSexplain}, the parameter $\alpha$ can speed up ($2>\alpha>1$) or slow down ($1>\alpha>0$) the convergence of the algorithm.
To illustrate such effect, the performances for different $\alpha\in\{0.5,1,1.5\}$ are shown in Fig.\:\ref{fig:IVAparamAlpha}, where the other parameters were set to $\mu_1=\mu_2=1$.
As expected, higher $\alpha$ achieved the scores at the final iteration with less number of iterations (note that the case for $\alpha=1.75$ is shown in Fig.\:\ref{fig:auxIVAvsProxIVA}).
For the Laplace IVA, the parameter $\alpha$ acted as a stretching factor of the horizontal axis.

\subsubsection{Effect of the Step-size Parameters $\mu_1$ and $\mu_2$}
\label{subsubsec:effectMu}

\begin{figure}[!t]
    \centering
    \includegraphics[width=0.98\columnwidth]{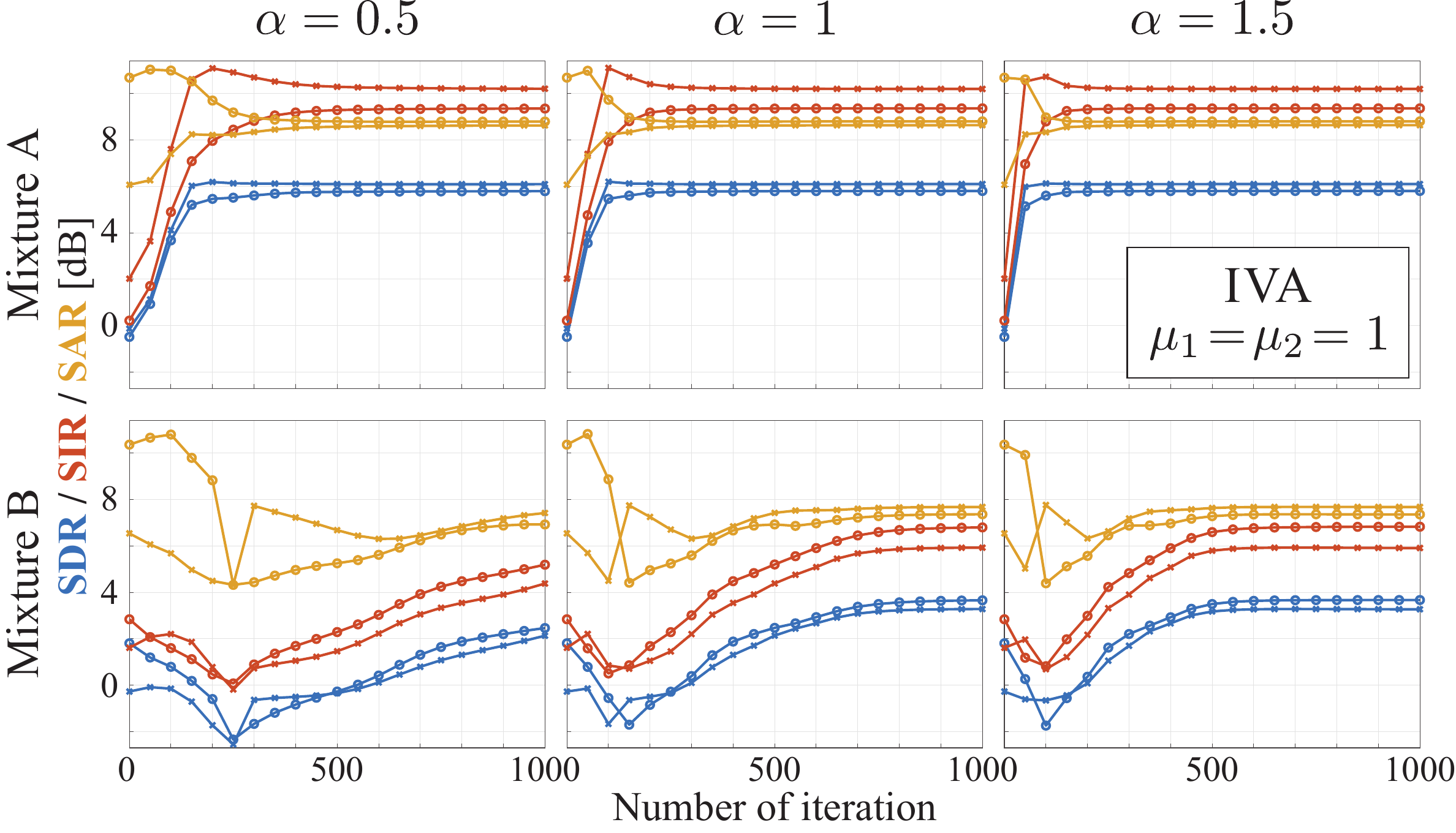}
    \caption{SDR/SIR/SAR of the signals separated from the mixtures in Fig.\:\ref{fig:micSpeakerSettingsSpeech}, where the parameter $\alpha$ was varied (Section \ref{subsubsec:effectAlpha}).}
    \label{fig:IVAparamAlpha}
\end{figure}

\begin{figure}[!t]
    \centering
    \includegraphics[width=0.98\columnwidth]{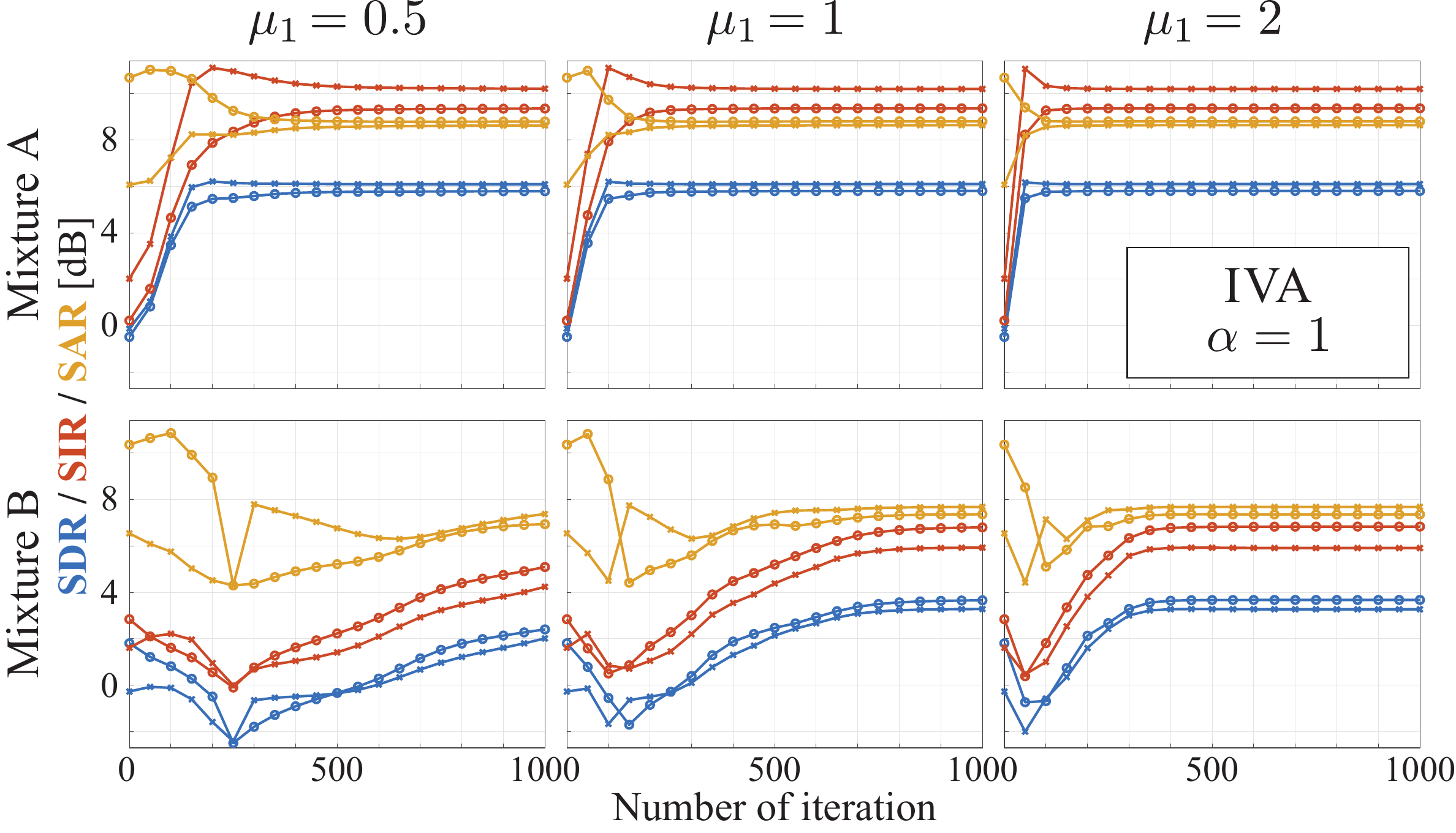}
    \caption{SDR/SIR/SAR of the signals separated from the mixtures in Fig.\:\ref{fig:micSpeakerSettingsSpeech}, where the parameter $\mu_1$ was varied, and $\mu_2=1/\mu_1$ (Section \ref{subsubsec:effectMu}).}
    \label{fig:IVAparamMu}
\end{figure}

As discussed in Section~\ref{sec:normalization}, the data normalization in Eq.~\eqref{eq:normalization} allows the choice $\mu_2=1/\mu_1$ for the step size, which comes from Eq.~\eqref{eq:muCond}.
Since $\mu_1$ and $\mu_2$ balance the effects of proximity operators ($\prox_{\mu_1 \mathcal{I}}$ and $ \prox_{\frac{1}{\mu_2}\mathcal{P}}$ in Algorithm~\ref{alg:PDS-BSS}), their choice can also affect the convergence.
By setting $\mu_2 = 1/\mu_1$ and $\alpha=1$, the performances of the PDS algorithm were investigated for $\mu_1\in\{0.5,1,2\}$ as illustrated in Fig.\:\ref{fig:IVAparamMu}.
From Figs.\:\ref{fig:IVAparamAlpha} and \ref{fig:IVAparamMu}, a specific choice of the parameters ($\mu_1$, $\mu_2$ and $\alpha$) seems to have little impact on the separation performance for the Laplace IVA if the number of iterations is sufficiently large.
In contrast, when the number of iterations is limited, these parameters may have some impact on the performance.

\subsection{Illustration of Basic Properties of HVA}
\label{subsec:HVAbasicProp}

As opposed to the Laplace IVA investigated in the previous subsection, HVA does not rely on a theoretical foundation but is heuristically defined by the time-frequency-masking function in Eq.~\eqref{eq:defHVAmaskFun}.
In such cases, theoretically developed criteria like Eq.~\eqref{eq:muCond} do not guarantee the convergence because their assumption is not satisfied anymore.
Therefore, the performance of HVA as well as its dependency on the algorithmic parameters must be investigated by experiments.
Here, the effects of step-size and relaxation parameters ($\mu_1$, $\mu_2$ and $\alpha$) are qualitatively presented.
The experimental conditions are the same as those in the previous subsection.
The parameters of the cepstrum thresholding were set to $\lambda=0.08$ and $\kappa=3$ based on the experimental results that will be presented in the next subsection.

\subsubsection{Effect of the Relaxation Parameter $\alpha$}
\label{subsubsec:effectAlphaHVA}

\begin{figure}[!t]
    \centering
    \includegraphics[width=0.98\columnwidth]{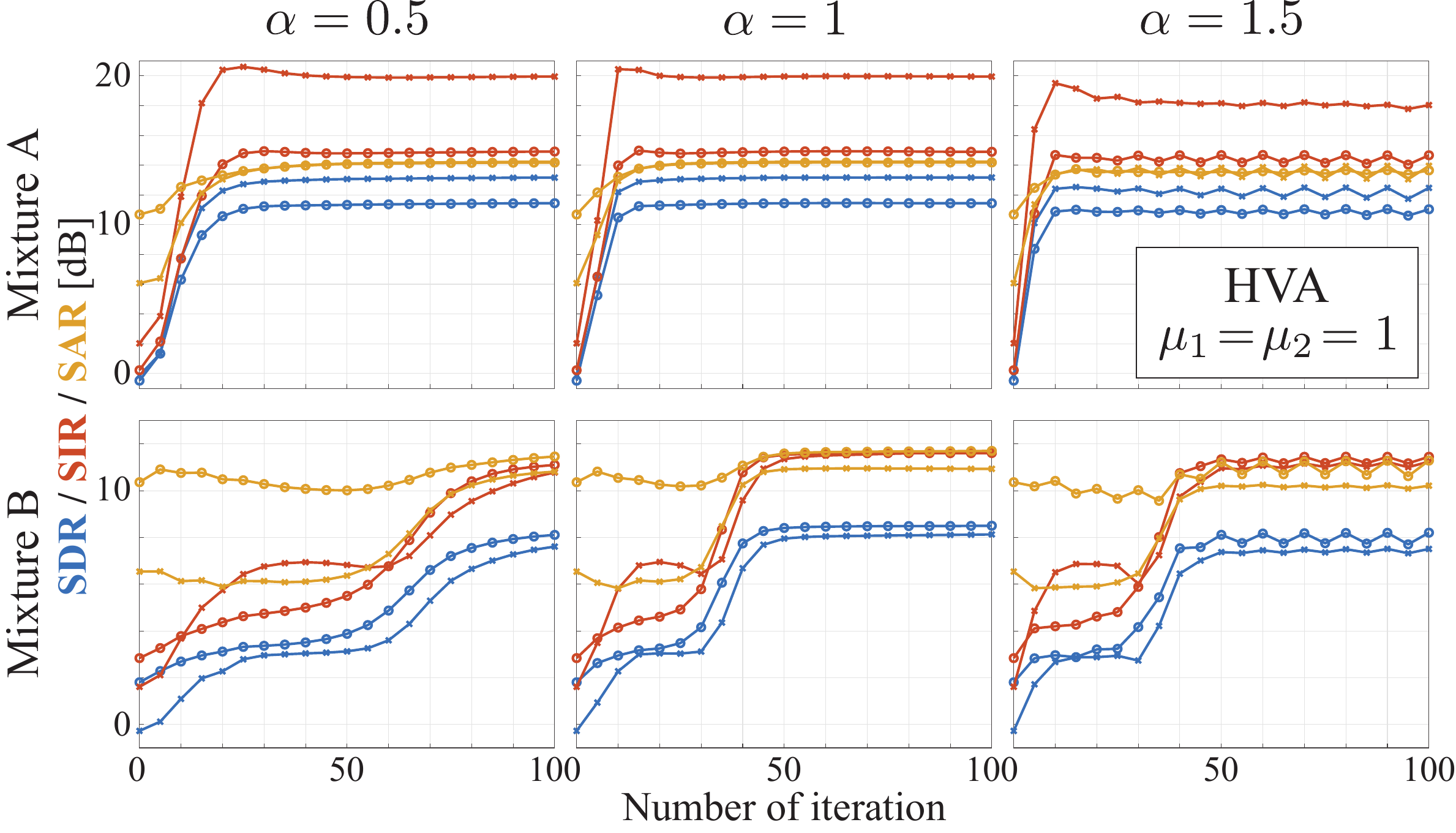}
    \caption{SDR/SIR/SAR of the signals separated from the mixtures in Fig.\:\ref{fig:micSpeakerSettingsSpeech}, where the parameter $\alpha$ was varied (Section \ref{subsubsec:effectAlphaHVA}).}
    \label{fig:HVAparamAlpha}
\end{figure}

At first, the effect of the relaxation parameter, $\alpha$, on the performance of HVA was investigated by fixing $\mu_1=\mu_2=1$.
The performances for $\alpha\in\{0.5,1,1.5\}$ are shown in Fig.\:\ref{fig:HVAparamAlpha} (note that the range of vertical and horizontal axes are greatly different from those in Figs.\:\ref{fig:IVAparamAlpha} and \ref{fig:IVAparamMu}).
Since the mask of HVA might not be stable compared to the proximity operator of a convex function, choice of $\alpha$ affected the performance at the final iteration.
For HVA, $\alpha=1$ seems a reasonable choice because it resulted in stable and fast improvement.
If a masking function is more unstable, a smaller $\alpha$ should be preferable for stabilizing the performance.

\begin{figure}[!t]
    \centering
    \includegraphics[width=0.98\columnwidth]{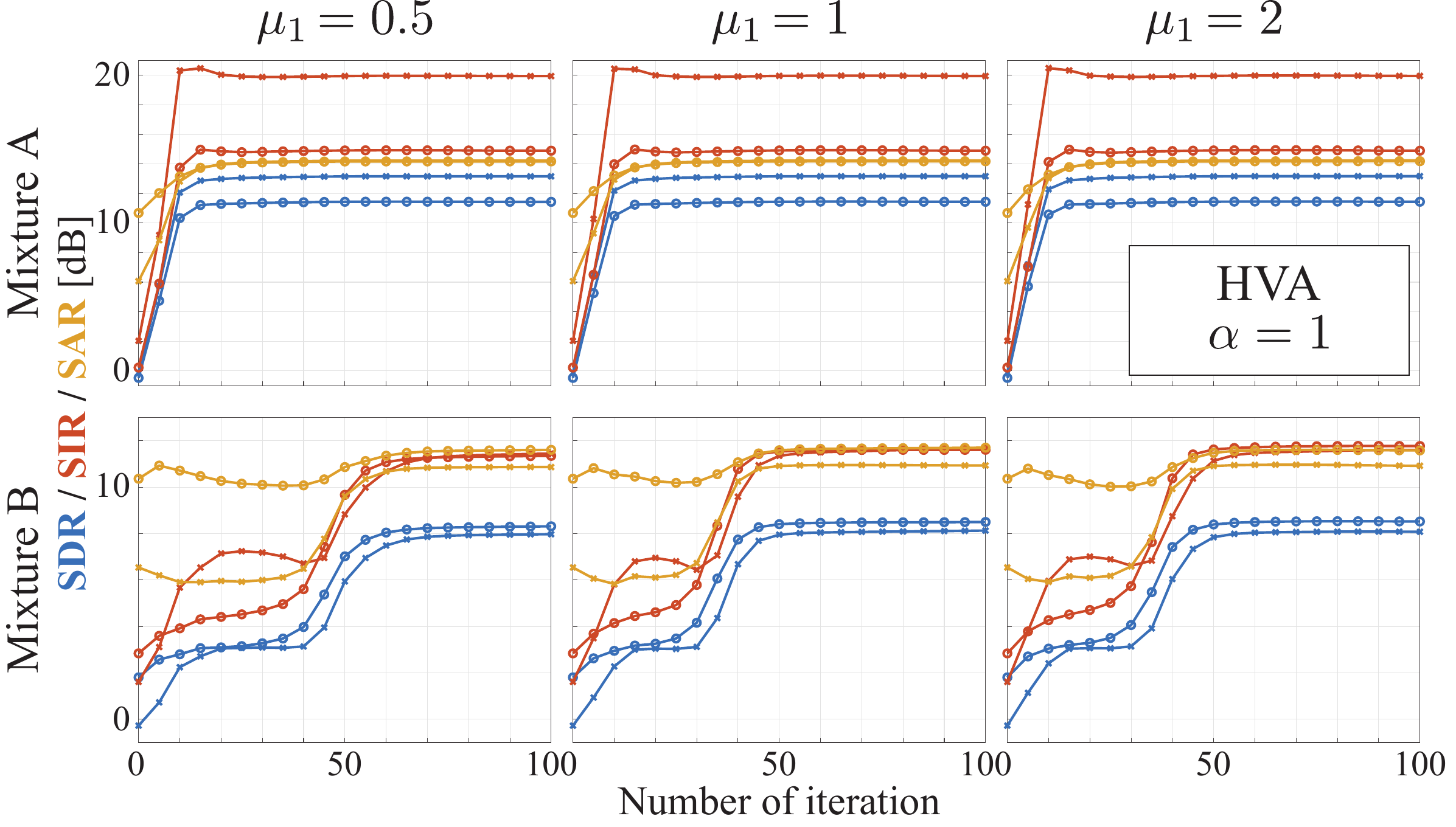}
    \caption{SDR/SIR/SAR of the signals separated from the mixtures in Fig.\:\ref{fig:micSpeakerSettingsSpeech}, where the parameter $\mu_1$ was varied, and $\mu_2=1/\mu_1$ (Section \ref{subsubsec:effectMuHVA}).}
    \label{fig:HVAparamMu}
\end{figure}

\subsubsection{Effect of the Step-size Parameters $\mu_1$ and $\mu_2$}
\label{subsubsec:effectMuHVA}

The effect of the step-size parameters $\mu_1$ and $\mu_2$ is also presented by fixing $\alpha=1$ as in Fig.\:\ref{fig:HVAparamMu}.
From the results, it can be seen that the effect of the choice of the step sizes on HVA is small.
This should be because the masking function of HVA in Eq.~\eqref{eq:defHVAmaskFun} is independent of $\mu_2$.
In general, such independence is not a favorable property of a masking function because the balance of the algorithm can collapse.
Even so, HVA was able to stably perform separation.


\subsubsection{Choice of the Parameters of HVA}

Based on the above experimental results, we suggest to choose the parameters of HVA as $\mu_1=\mu_2=\alpha=1$.
Since HVA is not so sensitive to $\mu_1$ and $\mu_2$, we suggest setting them to $1$ so that their multiplication can be omitted in the actual computation.
Also, the 7th and 8th lines of Algorithm \ref{alg:HVA} can be omitted in the actual computation when $\alpha=1$.

The other parameters of HVA, $\lambda$ and $\kappa$, will be investigated in the next subsection.
Overall, $\lambda=0.08$ seems a good choice, and $\kappa=3$ seems slightly better than $\kappa=1$ or $2$.




\subsection{Performance Evaluation of HVA using Speech Signals}
\label{subsec:speechExp}

For evaluating the performance of the proposed HVA, it is compared with the standard method, IVA, and the state-of-the-art method, ILRMA.
Here, we performed experiments using speech mixtures by improving the experiments in \cite{Kitamura2016}.

\subsubsection{Experimental Conditions}
\label{subsubsec:speechExpCond}

The database utilized in this experiment was a part of SiSEC (the underdetermined audio source separation task) \cite{SAraki2012_SiSEC}.
The BSS methods were evaluated using 2- and 3-channel speech mixtures.
For the 2-channel mixtures, 20 files (\texttt{male3}, \texttt{male4}, \texttt{female3}, and \texttt{male4}) contained in \texttt{dev1} and \texttt{dev2} were utilized.
They include live recordings (\texttt{liverec} containing ambient noise) with the reverberation time 130\,ms/250\,ms and the microphone spacing 1\,m/5\,cm.
The distances between the sources and the center of the microphone array is 1\,m.
For each situation, all possible pairs of the signals (out of 3 or 4) were selected to make the task determined ($N\!=\!M\!=\!2$).
As the result, 96 pairs for 20 situations were generated.%
\footnote{Note that, although \cite{Kitamura2016} utilized the same dataset, its experiment only contained 12 mixtures for 12 situations because the first two speech sources (out of 3) were selected for each situation.
In contrast, this paper utilized all signals included in each situation by selecting all possible pairs of the signals.}

Similarly, for the 3-channel mixtures, 8 files in \texttt{dev3} were utilized.
They include female/male speech with the reverberation time 130\,ms/380\,ms and the microphone spacing 50\,cm/5\,cm.
The distances between the sources and the center of the microphone array is 1\,m.
By selecting all possible 3-tuples from 4 signals to make the task determined ($N\!=\!M\!=\!3$), 32 mixtures for 8 situations were generated.

The half-overlapping 4096-point-long (256\,ms) Hann window was used for STFT as in \cite{Kitamura2016}.
All algorithms were iterated 200 times.
The number of bases of ILRMA for each source was set to 2, which is suitable for speech signals as shown in \cite{Kitamura2016}.
The parameters of HVA were set to $\alpha=\mu_1=\mu_2=1$, $\kappa\in\{1,2,3\}$, and $\lambda\in\{0.04,0.08,0.12,0.16,0.2\}$.
HVA without cepstrum thresholding ($\lambda=0$) was also tested.

\begin{figure}[!t]
    \centering
    \includegraphics[width=0.99\columnwidth]{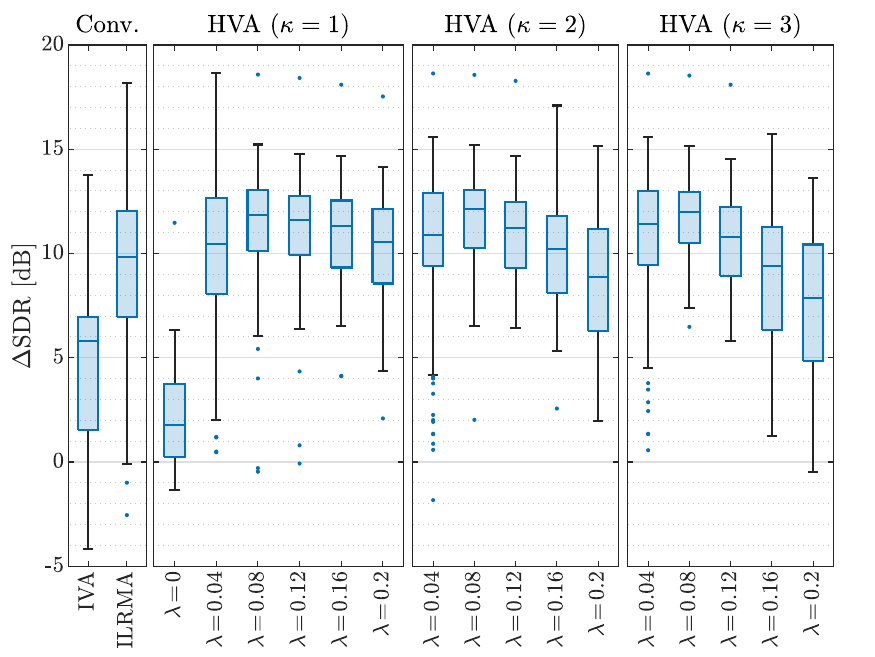}
    \caption{SDR improvements of the speech signals separated from 2-channel mixtures. The experimental settings are explained in Section \ref{subsubsec:speechExpCond}, where $\alpha=\mu_1=\mu_2=1$ and the number of iterations was 200. The central lines indicate the median, and the bottom and top edges of the box indicate the 25th and 75th percentiles, respectively.}
    \label{fig:boxSpeech2ch}
\end{figure}

\begin{figure}[!t]
    \centering
    \includegraphics[width=0.99\columnwidth]{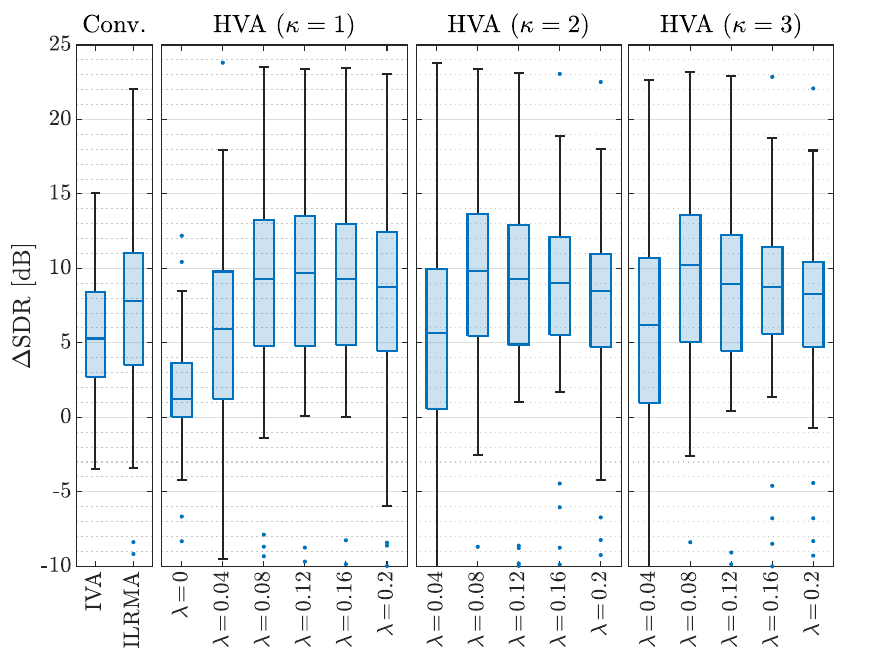}
    \caption{SDR improvements of the speech signals separated from 3-channel mixtures. The experimental settings are explained in Section \ref{subsubsec:speechExpCond}, where $\alpha=\mu_1=\mu_2=1$ and the number of iterations was 200.}
    \label{fig:boxSpeech3ch}
\end{figure}

\subsubsection{Results}

The experimental results for the 2- and 3-channel cases are summarized in Figs.\:\ref{fig:boxSpeech2ch} and \ref{fig:boxSpeech3ch}, respectively.
First of all, HVA without cepstrum thresholding ($\lambda=0$) could not perform separation.
As discussed in Section \ref{subsec:roleOfCepThresh}, this is because the permutation problem cannot be solved only by the non-separable mask.
This result indicates the importance of the cepstrum thresholding that simultaneously handles all frequency components.
From the figures, it can be seen that $\kappa$ has less impact than $\lambda$.
On average, $(\lambda,\kappa)=(0.08,3)$ seems to perform well, and therefore we will focus on HVA with $\lambda=0.08$ and $\kappa=3$.

The figures show that the proposed HVA outperformed IVA.
This result indicates that the harmonic structure can be a useful cue for the separation in determined BSS.
Compared to ILRMA, HVA achieved performance similar to ILRMA for the 3-channel case but outperformed it for the 2-channel case.
While ILRMA utilizes repetition of the spectral pattern with time as a cue for separation, HVA only focuses on the spectral pattern at each time segment independently, as illustrated in Fig.\:\ref{fig:maskExample}.
Since spectral patterns of speech signals widely vary with time, the low-rank structure (or repetitive pattern) of the magnitude spectrogram, assumed in ILRMA, may not effectively serve as a separation cue in this case.
In contrast, HVA is not hindered by such variation of signals because HVA considers time-independent information like IVA.
An audio example for each method can be found in \cite{KitamuraWeb}.

\begin{table}[!t]
	\caption{Music Signals for the 2-channel Mixtures \cite{SAraki2012_SiSEC}}
	\vspace{-4pt}
	\label{tab:miscSignals2ch}
	\centering
	\begin{tabular*}{0.95\columnwidth}{c @{\extracolsep{18pt}} c}
	\noalign{\hrule height .4mm}
		Song name & Source (1\texttt{/}2) \\\hline
		{bearlin-roads} & {acoustic\_guit\_main}\texttt{/}{vocals} \\
		{another\_dreamer-the\_ones\_we\_love} & {guitar}\texttt{/}{vocals} \\
		{fort\_minor-remember\_the\_name} & {violins\_synth}\texttt{/}{vocals} \\
		{ultimate\_nz\_tour} & {guitar}\texttt{/}{synth} \\
		{tamy-que\_pena\_tanto\_faz} & {guitar}\texttt{/}{vocals} \\
	\noalign{\hrule height .4mm}
	\end{tabular*}
\end{table}

\begin{table}[!t]
	\caption{Music Signals for the 3-channel Mixtures \cite{SAraki2012_SiSEC}}
	\vspace{-4pt}
	\label{tab:miscSignals3ch}
	\centering
	\begin{tabular*}{0.95\columnwidth}{c @{\extracolsep{\fill}} c}
	\noalign{\hrule height .4mm}
		Song name & Source (1\texttt{/}2\texttt{/}3) \\\hline
		{bearlin-roads} & {acoustic\_guit\_main}\texttt{/}{bass}\texttt{/}{vocals} \\
		{another\_dreamer-the\_ones\_we\_love} & {drums}\texttt{/}{guitar}\texttt{/}{vocals} \\
		{fort\_minor-remember\_the\_name} & {drums}\texttt{/}{violins\_synth}\texttt{/}{vocals} \\
		{ultimate\_nz\_tour} & {guitar}\texttt{/}{synth}\texttt{/}{vocals} \\
	\noalign{\hrule height .4mm}
	\end{tabular*}
\end{table}

\begin{figure}[!t]
    \centering
    \includegraphics[width=0.99\columnwidth]{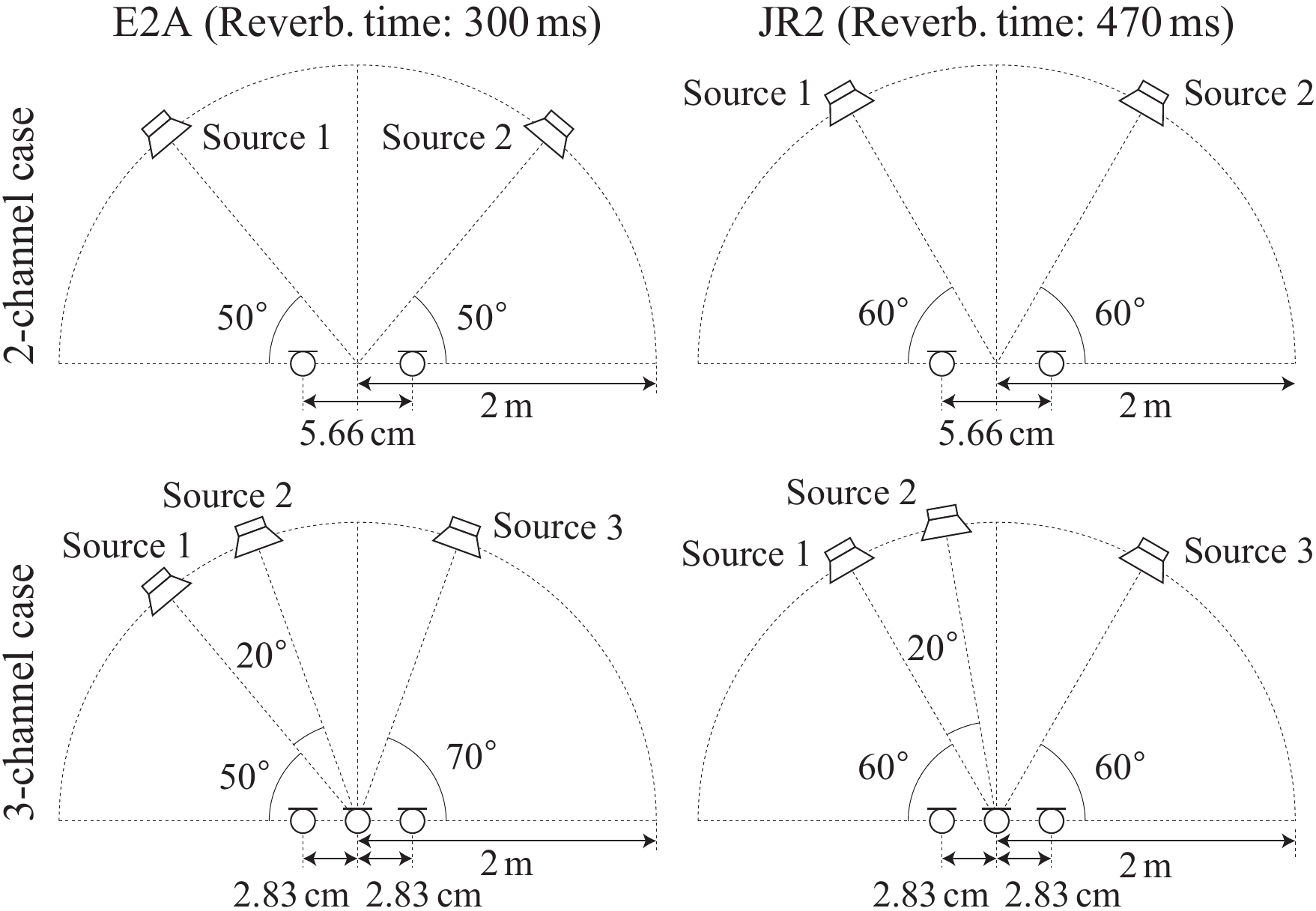}
    \caption{Recording conditions of the impulse responses (E2A and JR2 \cite{RWCP}) utilized in the experiment in Section~\ref{subsubsec:musicExpCond}.}
    \label{fig:micSpeakerSettingsMusic}
\end{figure}

\subsection{Performance Evaluation of HVA using Music Signals}
\label{subsec:musicExp}

Here, the proposed HVA was tested using music mixtures by following the experiments in  \cite{Kitamura2016}.

\subsubsection{Experimental Conditions}
\label{subsubsec:musicExpCond}

This experiment also used a part of SiSEC (the professionally produced music recordings) \cite{SAraki2012_SiSEC}.
The combinations of the source signals utilized in the 2- and 3-channel experiments are listed in Tables~\ref{tab:miscSignals2ch} and \ref{tab:miscSignals3ch}, respectively.
Since \texttt{tamy-que\_pena\_tanto\_faz} comprises only two sources (guitar and vocal), it was not included in the 3-channel case.
The 2- and 3-channel mixtures were produced by convolving the impulse response \texttt{E2A} or \texttt{JR2}, included in the RWCP database \cite{RWCP}, with each source. 
The recording conditions of these impulse responses are shown in Fig.\:\ref{fig:micSpeakerSettingsMusic}.

As in \cite{Kitamura2016}, the 3/4-overlapping 8192-point-long Hann window (512\,ms) was used for STFT, and the number of bases of ILRMA for each source was set to 30.
The other settings of the algorithmic parameters were the same as those in the previous experiment using speech signals.

\begin{figure}[!t]
    \centering
    \includegraphics[width=0.99\columnwidth]{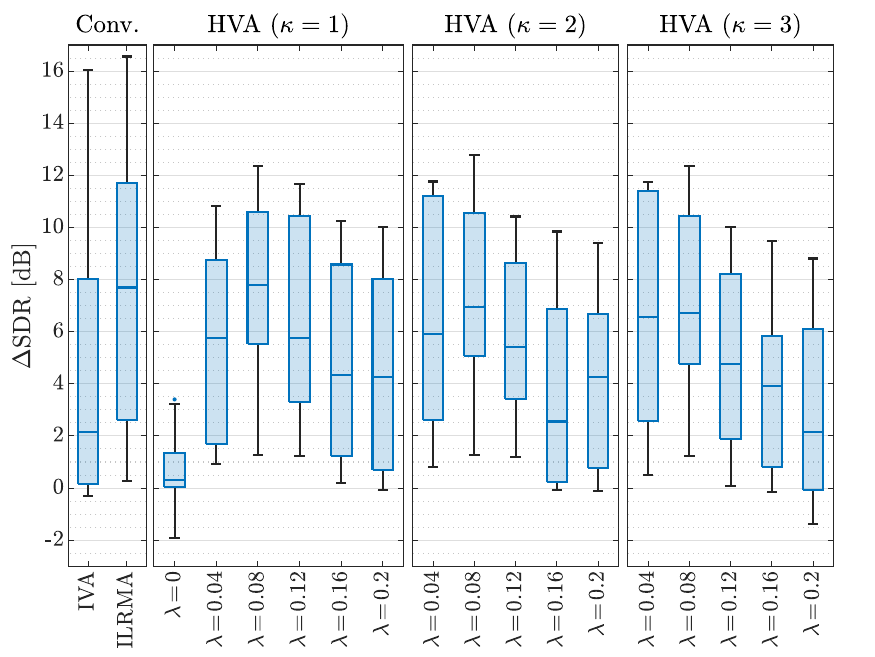}
    \caption{SDR improvements of the music signals separated from 2-channel mixtures. The experimental settings are explained in Section \ref{subsubsec:musicExpCond}, where $\alpha=\mu_1=\mu_2=1$ and the number of iterations was 200.}
    \label{fig:boxMusic2ch}
\end{figure}

\begin{figure}[!t]
    \centering
    \includegraphics[width=0.99\columnwidth]{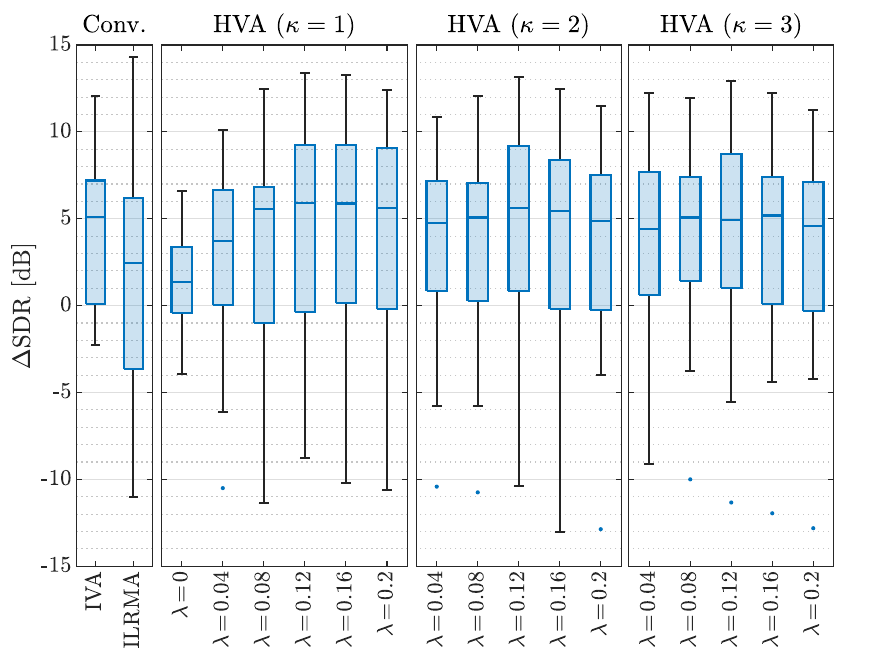}
    \caption{SDR improvements of the music signals separated from 3-channel mixtures. The experimental settings are explained in Section \ref{subsubsec:musicExpCond}, where $\alpha=\mu_1=\mu_2=1$ and the number of iterations was 200.}
    \label{fig:boxMusic3ch}
\end{figure}

\subsubsection{Results}

The experimental results for the 2- and 3-channel cases are summarized in Figs.\:\ref{fig:boxMusic2ch} and \ref{fig:boxMusic3ch}, respectively.
From the figures, it can be seen that the proposed HVA was comparable to the other methods.
Since the 2-channel mixtures comprise harmonic signals as the sources (see Table~\ref{tab:miscSignals2ch}), HVA should have been able to effectively model the harmonic structure through cepstrum.
Therefore, even though ILRMA's ability of modeling repetitive spectral patterns is advantageous for music signals, HVA with $\lambda=0.08$ was able to perform similarly.

For the 3-channel case, IVA performed better than ILRMA.
One reason for this result should be the presence of drums in the sources (see Table~\ref{tab:miscSignals3ch}). 
ILRMA can handle such percussive sources by devoting some of the bases, but HVA does not have a mechanism for explicitly handle percussive sources.
Even so, HVA was able to perform comparably in this case, too.
Note that, as the masking-based BSS framework can simultaneously utilize multiple criteria by combining multiple masks, HVA has a potential of improvement by incorporating other time-frequency masks targeting at the specific structure of the source signals as examined in \cite{Oyabu2021ICASSP}.

\begin{table}[!t]
	\caption{Computational Time per Iteration (w/ Core i7-8700)}
	\vspace{-4pt}
	\label{tab:computTime}
	\centering
	\begin{tabular*}{0.95\columnwidth}{c|cc}
	\noalign{\hrule height .4mm}
		 & Speech (Section \ref{subsec:speechExp}) & Music (Section \ref{subsec:musicExp}) \\\hline
		auxIVA & 47.6 ms & 106.0 ms \\
		ILRMA & 116.1 ms & 271.2 ms \\
		HVA (Proposed) & 60.8 ms & 164.3 ms \\
	\noalign{\hrule height .4mm}
	\end{tabular*}
\end{table}

\subsection{Computational Efficiency}

The computational time of the algorithms depend on the settings such as the window length and the number of channels.
Table \ref{tab:computTime} shows the computational time per iteration for 2-channel speech and music separation (Section \ref{subsec:speechExp} and \ref{subsec:musicExp}, respectively) as an example.
Here, the computational time for the speech and music cases were different because the numbers of time segments and frequencies were different.
The computational time of HVA was more than that of IVA because HVA involves additional computation of logarithm, the (frequency-directional) Fourier transform pair, cosine, and exponential.
ILRMA required more time for computation because it treats the demixing filters as well as all time-frequency bins as the optimization variables. 
As in Table \ref{tab:computTime}, the proposed HVA tends to be more efficient than ILRMA but less efficient than IVA.

Note that Table \ref{tab:computTime} shows computational time \textit{per iteration}, and the total time depends on the number of iterations.
The required number of iterations depends on the situations.
For example, as shown in Figs.~\ref{fig:HVAparamAlpha} and \ref{fig:HVAparamMu}, HVA seems to require 20 and 50 iterations for Mixture A and B, respectively, with the setting explained in Section \ref{subsec:basicProp}.
Since HVA is efficient than ILRMA for each iteration and requires relatively small number of iterations (see Fig.\:\ref{fig:maskEachIter} for an example of the rapid evolution of the mask of HVA), the computational time of HVA is usually less than ILRMA.

\subsection{Limitations of HVA}

Here, some limitations of HVA are discussed.
First, since the mask-generating function of HVA targets only harmonic signals, HVA cannot directly handle signals that do not exhibit the harmonic structure (e.g., white noise).
The mask must be modified to enhance the non-harmonic signals in that case.
Second, the assumed model in HVA (also in IVA and ILRMA) [Eq.~\eqref{eq:mixingProcess}] does not consider additional noise and/or time variation of the mixing system.
To handle such cases, the algorithm and mask must be modified accordingly.
Third, the mask of HVA cannot be used as a post filter as in \cite{modelBasedIVA} because the mask is not intended to separate signals.

Although HVA has these limitations, some of them can be easily resolved because defining another mask-generating function is painless.
Note that the ideas behind the mask of HVA can be solely used in other signal processing methods that target harmonic signals.
Such extensions and applications of the proposed method and ideas are left as the future works.

\section{Conclusions}
\label{sec:conclusion}

In this paper, the novel BSS method termed HVA was proposed.
By modeling the harmonic structure via cepstrum analysis, HVA achieved the performance comparable to the state-of-the-art method, ILRMA, with less computational effort.
To realize HVA, the general BSS algorithm based on time-frequency masking was presented.
Since it allows any mask for enhancing the source signals, improving HVA as well as investigating a completely new BSS method should be easy.
The future works include an extension of HVA by data-adaptation (instead of using the fixed masking function), online extension of the algorithm using adaptive techniques \cite{OnoAdaptivePDS}, and investigation of combination of the masking-based algorithm and the existing source enhancement techniques.

\ifCLASSOPTIONcaptionsoff
  \newpage
\fi



\bibliographystyle{IEEEtran}
\bibliography{IEEEabrv,refs}
%



%


\begin{IEEEbiography}{Kohei Yatabe}
received his B.E., M.E., and Ph.D. degrees from Waseda University in 2012, 2014, and 2017, respectively. He is currently an assistant professor of the Department of Intermedia Art and Science, Waseda University. His research interests include optical measurement of airborne sound.
\end{IEEEbiography}


\begin{IEEEbiography}{Daichi Kitamura}
received the Ph.D. degree from SOKENDAI, Hayama, Japan. He joined The University of Tokyo in 2017 as a Research Associate, and he moved to National Institute of Technology, Kagawa Collage as an Assistant Professor in 2018. His research interests include audio source separation, statistical signal processing, and machine learning. He was the recipient of the Awaya Prize Young Researcher Award from The Acoustical Society of Japan (ASJ) in 2015, Ikushi Prize from Japan Society for the Promotion of Science in 2017, Itakura Prize Innovative Young Researcher Award from ASJ in 2018, and Young Author Best Paper Award from IEEE Signal Processing Society in 2019.
\end{IEEEbiography}


\vfill


\end{document}